\documentclass[prd,aps,reprint,nofootinbib,superscriptaddress]{revtex4-2}

\usepackage{amsmath,amsfonts,amssymb,amsthm,
            mathtools,
            esint,
            hyperref,
            url,
            mathrsfs,
            graphicx,
            stmaryrd}

\usepackage[nodisplayskipstretch]{setspace}
\usepackage[utf8]{inputenc}
\usepackage[T1]{fontenc}
\usepackage{lmodern}
\usepackage[version=4]{mhchem}
\usepackage{stmaryrd}
\usepackage{amsbsy}
\usepackage{latexsym}
\usepackage{comment}
\usepackage{natbib}
\usepackage{bm}
\usepackage{subfigure}
\usepackage{color}
\usepackage{wasysym}
\usepackage{mathbbol}
\usepackage{microtype}
\usepackage{bigints}
\allowdisplaybreaks
\usepackage[normalem]{ulem}
\usepackage[dvipsnames]{xcolor}
\usepackage{multirow}
\usepackage{physics}
\usepackage[export]{adjustbox}
\usepackage{booktabs} 

\hypersetup{
	colorlinks=true,       
	linkcolor=blue,          
	citecolor=blue,        
	urlcolor=blue           
}

\newcommand{\newtext}[1]{\textcolor{NavyBlue}{#1}}

\usepackage{hyperref}
\begin{document}

\title{Anomaly quenching and dynamical cooling of Hawking evaporation in Horndeski gravity}

\author{Amogh Srivastav}
\email{23b1826@iitb.ac.in}

\author{S.~Shankaranarayanan}
\email{shanki@iitb.ac.in}

\affiliation{Department of Physics, Indian Institute of Technology Bombay,
             Mumbai 400076, India}

\begin{abstract}
We investigate the semiclassical Hawking evaporation of a two-dimensional Callan-Giddings-Harvey-Strominger (CGHS) black hole perturbed by first-order Horndeski scalar-tensor couplings. By evaluating the exact scalar curvature invariants, we demonstrate that coordinate singularities near the apparent horizon parametrize a physical deformation of the surface gravity while preserving a strictly regular event horizon. In the linear-order $\gamma_3 \Phi X$ coupling sector, the surface gravity and correspondingly the Hawking temperature decreases dynamically as the black hole loses mass. Extrapolating this cooling effect via a resummed temperature equation suggests that evaporation halts, potentially leaving behind a stable, macroscopic cold remnant at the boundary of perturbative control. Utilizing both the conformal trace anomaly and the Robinson-Wilczek gravitational anomaly, we establish that the asymptotic radiation flux identically reflects this geometric shift. Consequently, the fine-grained entanglement entropy of the Hawking radiation departs from linear growth and transitions to an asymptotic logarithmic regime. This indicates that quantum information may be permanently retained within the remnant, altering the standard semiclassical Page curve without requiring a breakdown of horizon regularity.
\end{abstract}

\maketitle

\section{Introduction}
The black hole (BH) information paradox
\cite{Hawking:1975vcx,Hawking:1976ra,Calmet:2022swf} remains the
ultimate stress test for any candidate theory of quantum gravity.  At
the heart of the paradox lies a breakdown of semiclassical spacetime
during the final stages of evaporation: as a BH of mass $M$ radiates, its Hawking temperature 
$$T_H = (\hbar c^3/Gk_B)(8\pi M)^{-1}$$
diverges, and the Kretschmann invariant
$R_{\mu\nu\rho\sigma}R^{\mu\nu\rho\sigma} \propto (GM)^{-4}$ signals
the collapse of classical general
relativity~\cite{Shankaranarayanan:2022wbx,Mandal:2025xuc}.  At these
extreme Planckian curvatures, higher-order derivative corrections cease
to be negligible perturbative adjustments; they become the dominant
dynamical operators governing the fate of the geometry.  Incorporating
such corrections into a unitary quantum field theory without introducing
the Ostrogradsky instability remains a central theoretical
hurdle~\cite{Ostrogradsky:1850fid,Mandal:2025xuc}.

A highly productive arena for isolating this physics is 2-D
dilaton gravity.  Since the $s$-wave reduction of 4-D
spacetime captures roughly $90\%$ of the total emitted Hawking flux
\cite{Page:1976df,Sanchez:1977si}, $(1{+}1)$-dimensional models
preserve the causal and thermodynamic core of the evaporation problem
while remaining analytically tractable.
%
A comprehensive study by Martinec~\cite{Martinec:1996bh} emphasized that these 2-D models provide a controlled setting for examining Hawking radiation, semiclassical backreaction, and the quantum consistency of BH evaporation.
Among these, the
Callan-Giddings-Harvey-Strominger (CGHS) model \cite{Callan:1992rs}
stands as the paradigmatic workhorse: it admits exact classical BH
solutions, yields a thermal flux, and has served as the primary laboratory for studying quantum backreaction for over three decades
\cite{Russo:1992ax,Russo:1992ht,Harvey:1992xk,Strominger:1995xw}.
However, the canonical CGHS action possesses no native, systematic
procedure for incorporating higher-curvature corrections, leaving the terminal phase of the Page
curve~\cite{Page:1993df,Page:1993wv} fundamentally unresolved.

Recently, Mandal et al.~\cite{Mandal:2023kpu} proved that the exact CGHS action arises via the spherical reduction of 4D Horndeski gravity, the most general second-order scalar-tensor theory whose Euler-Lagrange equations remain strictly second-order in the metric $g^{(4D)}_{\mu\nu}$ and the scalar field $\Phi$~\cite{Horndeski:1974wa}. They showed that the Horndeski framework thus opens a rigorous, ghost-free gateway to imprint trans-Planckian corrections onto the CGHS geometry via the higher-order coupling functions $G_4(\Phi, X)$ and $G_5(\Phi, X)$ (where $X = -\nabla^\mu \Phi \nabla_\mu \Phi/2$) in Horndeski gravity.  However, Ref.~\cite{Mandal:2023kpu} restricted its analysis to the $\partial_X G_{4} = \partial_X G_{5} = 0$ subspace to avoid the integration-by-parts (IBP) ambiguities inherent to dimensional reduction. This restriction obscures the full dynamical impact of
higher-derivative operators on the quantum stress tensor.

In this work, we lift this constraint by formulating the exact 2-D
dilaton gravity theory across all three independent higher-derivative sectors ($\partial_X G_{4}, G_5, \partial_X G_5$).  By deriving the exact 2-D commutator constraints required to tame the IBP remainders, we integrate the dynamical field equations in null gauge over a collapsing shockwave background.  We establish that the interplay of the Horndeski coupling sectors dynamically quenches the Robinson-Wilczek gravitational anomaly, forcing the evaporation to halt at a macroscopic cold remnant and driving
the semiclassical Page curve into an asymptotic logarithmic saturation.

We first establish the linear-order corrections to BH thermodynamics dictated by the Horndeski terms. We show that the Robinson-Wilczek gravitational anomaly yields a systematic cooling effect, governed by the evolving dimensionless parameter $\gamma_3\lambda^5/M^3$. As the BH evaporates and its mass $M$ decreases, this parameter approaches unity, signaling the boundary of our perturbative effective field theory (EFT) expansion. To investigate the thermodynamic trajectory near this threshold, we introduce a resummed temperature profile:
$$T_H(M) = \frac{\lambda}{2\pi}\left(1-\frac{M_{\rm r}^3}{M^3}\right).$$
Here, $\gamma_3$ is the coefficient of the linear-order $G_3=\gamma_3\Phi X$ Horndeski coupling introduced in Sec.~III, and $M_{\rm r}\equiv(2\gamma_3\lambda^5/3)^{1/3}$ is the resulting critical remnant mass, such that the dimensionless EFT expansion parameter is exactly $(M_{\rm r}/M)^3$. While this closed-form expression represents an extrapolation to the boundary of the perturbative regime rather than a definitive non-perturbative proof, it rigorously exponentiates the linear-order coordinate shift to provide a thermodynamically consistent model for the endpoint regime. This framework demonstrates that the anomalous corrections dynamically quench the evaporation process as $M \rightarrow M_{\rm r}$, leading to the formation of a stable, macroscopic cold remnant at the limit of perturbative validity.


The remainder of this paper is organized as follows. In Sec.~II, we introduce the model by dimensionally reducing the four-dimensional Horndeski action to the two-dimensional modified CGHS framework. Section~III details the exact field equations in conformal null gauge and extracts the leading-order dynamical solutions, identifying a strong-coupling divergence at the classical apparent horizon. In Sec.~IV, we calculate the scalar curvature invariants to prove this divergence is a coordinate artifact and isolate the strictly regular, physically shifted event horizon. Section~V constructs the resummed temperature profile and analyzes the dynamical cooling that leads to the formation of the macroscopic remnant. Section~VI computes the exact semiclassical stress tensor and the chiral gravitational anomaly, proving that the macroscopic luminosity shift matches the microscopic anomaly quenching. Section~VII evaluates the fine-grained entanglement entropy of the Hawking radiation and establishes its late-time logarithmic saturation. Finally, we summarize our physical conclusions in Sec.~VIII. Exhaustive derivations of the field equations, coordinate transformations, anomaly integrals, and entropy profiles are collected in the Appendices.

Throughout this work, we adopt the following conventions. We work in natural units where $\hbar = c = k_B = 1$. Quantities intrinsically defined on the 4-D spacetime, such as the full metric and Ricci scalar, are explicitly denoted with a superscript $(4D)$ (e.g., $g_{\mu\nu}^{(4D)}$ and $R^{(4D)}$). Greek indices ($\mu, \nu, \dots$) run over the four-dimensional spacetime coordinates from $0$ to $3$, and we employ the mostly-plus metric signature $(-, +, +, +)$. The dimensionally reduced two-dimensional metric and its corresponding Ricci scalar are denoted simply by $g_{ab}$ and $R$, with Latin indices ($a, b, \dots$) running over the 2-D spacetime coordinates $0$ and $1$, carrying the induced signature $(-, +)$.

\section{4D Horndeski to 2D Dilaton Gravity}
Consider the 4D Horndeski action~\cite{Kobayashi:2019hrl}
\begin{equation}
S_{\rm Horndeski} = \int d^4x \sqrt{-g^{(4D)}} \sum_{i=2}^5
(\mathcal{L}_{G_i} + \mathcal{L}_{G_{i,X}}) \, .
\label{eq:Horndeski}
\end{equation}
where $G_{i,X}\equiv\partial G_i/\partial X$. (Appendix~\ref{app:action} contains all the terms in the action.)  We impose the standard spherically symmetric reduction ansatz upon the 4-D line element and scalar field ($\Phi$):
\begin{equation}
ds^2_{(4D)} = g_{ab}(x^c)dx^a dx^b + \rho^2(x^c)d\Omega^2,
\quad \Phi = \Phi(x^c),
\label{eq:ansatz}
\end{equation}
where $a,b\in\{0,1\}$ span the 2-D Lorentzian manifold $\mathcal{M}_2$ with signature $(-,+)$, $\rho(x^a)$ is the dynamical areal radius, and $d\Omega^2$ is the unit two-sphere.  Integrating out the angular coordinates yields the universal volume factor $\int d\Omega = 4\pi$. To organize the resulting 2-D action $S^{(2D)} = \int d^2x\sqrt{-g}\,\mathcal{L}_{\rm 2D}$, we define the fundamental first-order kinematic inner products on $\mathcal{M}_2$:
\begin{equation}
X = -\tfrac{1}{2}\nabla^a\Phi\nabla_a\Phi,\quad
Y = \nabla^a\rho\nabla_a\rho,\quad
Z = \nabla^a\Phi\nabla_a\rho
\label{eq:kin_1}
\end{equation}
alongside the second-order Hessian traces (details in  Appendix~\ref{app:action}):
\begin{multline}
\mathcal{ I} = \nabla_a\nabla_b\Phi\nabla^a\nabla^b\Phi,\quad
W = \nabla_a\nabla_b\Phi\nabla^a\nabla^b\rho, \\
\mathcal{C} = \nabla_a^{\ b}\Phi\nabla_b^{\ c}\Phi\nabla_c^{\ a}\Phi.
\label{eq:kin_2}
\end{multline}
In contrast to Ref.~\cite{Mandal:2023kpu}, which circumvented
dimensional reduction ambiguities by imposing $G_{4,X}=G_{5,X}=0$ as an \emph{a priori} assumption at the 4-D level, we execute the reduction of the Horndeski action \eqref{eq:Horndeski} to uncover its underlying 2-D constraint structure.  The dimensionally reduced Galileon sectors naturally project onto 2-D operators containing the problematic kinetic combination $(\Box\Phi)^2 - \mathcal{ I}$ (where $\Box$ is the 2-D d'Alembertian operator). To systematically organize these terms, we deploy the 2-D Ricci commutator identity,
$$(\Box\Phi)^2 - \mathcal{ I} = -R\,X + \nabla_a J^a,$$
where the current is defined as $J^a = (\Box\Phi)\nabla^a\Phi -
\nabla^a\nabla^b\Phi\,\nabla_b\Phi$ and $R$ is the 2-D Ricci scalar.

Applying this identity elegantly shifts the higher-derivative operators into pure geometric couplings proportional to $R X$.  However, integrating the divergence $\nabla_a J^a$ by parts generates a cascade of surface remainders proportional to $G_{4,X},G_{4,XX},G_{4,X\Phi}$ and $G_{5,X},G_{5,XX}$.  To preserve strict ghost-freedom, these remainders must be systematically removed at the source.  This requirement imposes
the exact operational selection rule:
\begin{equation}
G_{4,X} = G_{5,X} = 0 \;\iff\;
G_4 = G_4(\Phi),\ G_5 = G_5(\Phi).
\label{eq:selection}
\end{equation}
Consequently, the truncation previously treated as an \emph{ad hoc}
simplification in the literature is revealed here as a rigorous
topological necessity for 2-D stability.  The surviving, fully
consistent 2-D Lagrangian density is exactly given by (details in  Appendix~\ref{app:action}):
\begin{equation}
{\cal L}_{(2D)} = L_{G_2}(\Phi,X) + L_{G_3}(\Phi,X)
               + L_{G_4}(\Phi)   + L_{G_5}(\Phi)\,,
\label{eq:S2}
\end{equation}
leaving the functions $G_2(\Phi,X)$ and $G_3(\Phi,X)$ completely
unrestricted, alongside the purely field-dependent couplings
$G_4(\Phi)$ and $G_5(\Phi)$.

Using the Brans-Dicke mapping $\Phi=e^{2\varphi}$, $\rho=\Lambda
e^{-2\varphi}$, and $G_4=\Phi$~\cite{Mandal:2023kpu}, the choices $\Omega=-6/\Phi$ and $V=2\Phi^3/\Lambda^2-4\lambda^2\Phi$ identically recover the classical CGHS action: 
$\mathcal{L}_{\rm CGHS} = \Lambda^2
e^{-2\varphi}[R+4(\nabla\varphi)^2+4\lambda^2]$~\cite{Mandal:2023kpu}.  
Higher-derivative trans-Planckian corrections
are thus governed entirely by the functional form of  $G_3(\Phi,
X_\Phi)$ and $G_5(\Phi)$. {The complete intermediate steps of this reduction, the explicit 4-D Horndeski densities, their spherically reduced 2-D counterparts, and the Brans-Dicke gauge-fixing that isolates the CGHS target space, including an explicit check that no factor of $\rho^2$ is dropped in the $G_2$ sector are collected in Appendix~\ref{app:action}.}

\section{Dynamical equations and Strong-Coupling Horizon}
\label{sec:Dynamics}

We model BH as forming from the collapse of an infalling matter shockwave onto flat space, which produces the classical, static CGHS geometry outside the shockwave. The Horndeski couplings $G_3=\gamma_3\Phi X$ and $G_5=\gamma_5\Phi$ are then switched on as perturbations on this static background to track the resulting semiclassical evaporation. To integrate the semiclassical collapse, we rewrite the exact field equations into conformal null gauge: 
$$ds^2=-e^{2\omega}dx^+dx^-.$$ 
This yields the dilaton equation $\mathcal E_\varphi=0$, the conformal constraint $\mathcal E_\omega=0$, and the chiral constraint $\mathcal E_{++}=0$. Their full forms for general $G_3(\Phi,X)$ and $G_5(\Phi)$ are detailed in Appendix~\ref{app:eom}. 

We define the Kruskal-type variables $u\equiv e^{-2\varphi}$ and $v\equiv e^{-2\omega}$. To solve the field equations perturbatively, we specialize to the static, spherically-collapsed background relevant for an adiabatically evaporating BH by introducing \emph{static Kruskal coordinates} $X^\pm$. It is convenient to rewrite the product $\chi\equiv X^+X^-$ for the areal-type radial variable $q$:
$$q \equiv \frac{M}{\lambda}-\lambda^2 X^+X^-.$$
The unperturbed CGHS geometry is perfectly static in this coordinate. On this background, the fields degenerate to $u_0=v_0=q$, mapping spatial infinity to $q\to\infty$ and the classical event horizon ($X^+X^-=0$) to $q\to M/\lambda$. Because $q$ is the unique static slice coordinate in which the unperturbed dilaton and conformal factor coincide identically, a nonzero ratio $\alpha=u/v \neq 1$ at any order is an \emph{unambiguous, gauge-invariant signature due to the Horndeski corrections}.

Turning on the Horndeski coupling constants, $G_3=\gamma_3\Phi X$ and $G_5=\gamma_5\Phi$, backreacts on this geometry. To extract the exact leading-order dynamics, we perform a perturbative expansion in the couplings $\gamma_i$, which carry dimensions of (length)$^2$ so that the combinations $\gamma_3\lambda^5/M^3$ and $\gamma_5\lambda^3/\Lambda^2M^3$ controlling the expansion are dimensionless. This captures the geometric deformations via the shift functions $A_i(q)$ and $W_i(q)$ (details in Appendix~\ref{app:eom}):
$$u = \alpha v = q + \gamma_i[A_i(q)q + W_i(q)] + \mathcal{O}(\gamma_i^2).$$
Substituting this ansatz into the exact field equations (using the static-slice reduction $\partial_+\partial_-F = F' + \chi F''$ of Eq.~\eqref{eq:app_kruskal_static}, which converts every null derivative into a $q$-derivative; see Appendix~\ref{app:static_q}) linearizes the system into two coupled, second-order ordinary differential equations for $A_i(q)$ and $W_i(q)$ in each sector.

For both the $G_3$ and $G_5$ sectors, the system can be decoupled. With the normalization of the linearized equations fixed as in Appendix~\ref{app:linearized}, the linear combination $\mathcal{E}_\varphi^{(1)} + q\,\mathcal{E}_\omega^{(1)}$ in the $G_3$ sector, and $\mathcal{E}_\varphi^{(1)} + \mathcal{E}_\omega^{(1)}$ in the $G_5$ sector, strictly eliminates both $W_i''$ and $W_i'$ simultaneously. Dividing the resulting constraint by the common overall factor of $4q^6$ ($G_3$ sector) or $4q^5$ ($G_5$ sector) yields a universal master equation for the geometric ratio correction $A_i(q)$:
\begin{equation}
(\lambda q-M)A_i''(q)+\lambda A_i'(q)=\mathcal{S}_i(q),
\label{eq:Masterequation}
\end{equation}
where primes denote derivatives with respect to $q$, and the sources $\mathcal{S}_i(q)$ are algebraic rational functions (given explicitly in Appendix~\ref{app:linearized}). The homogeneous solution to this master equation is 
\begin{equation}
A_i^{(H)}(q) = C_i + D_i\ln(\lambda q-M) \, , 
\label{eq:Homogeneous}
\end{equation}
where $C_i$ and $D_i$ are integration constants. Demanding asymptotic flatness uniquely fixes these constants: bounding the geometry at spatial infinity ($q\to\infty$) strictly requires $D_i = C_i=0$. 

With the boundary conditions fixed, the inhomogeneous master equation is integrated exactly via variation of parameters. The profile $W_i(q)$ is subsequently recovered algebraically by substituting $A_i(q)$ back into the linearized constraint equations. The closed-form solutions are given in Appendix~\ref{app:closed_form}. 

The asymptotic behaviors of these exact solutions dictate the structural fate of the geometry. At spatial infinity, the corrections decay rapidly as $A_3\propto q^{-3}$ and $A_5\propto q^{-1}$, confirming that the Horndeski operators strictly preserve the asymptotic linear-dilaton structure. However, approaching the classical horizon ($q\to M/\lambda$), the geometric corrections exhibit a logarithmic divergence:
$$A_3 \sim -\frac{2\lambda^5}{3M^3}\ln\epsilon,\quad
A_5 \sim -\frac{4\lambda^3}{M} \left(1 -\frac{1}{12\Lambda^2M^2}
  \right)\ln\epsilon,$$
where $\epsilon = q - M/\lambda$ is the infinitesimal proper distance. This divergence indicates that both the $G_3$ and $G_5$ corrections become strongly coupled near the apparent horizon. The breakdown of the perturbative parameter $\alpha=u/v\to1$ fundamentally disrupts the classical coordinate chart at the horizon, signaling the physical mechanism that alters standard evaporation and forces the freeze-out of a macroscopic remnant.

\section{Curvature Invariants and Horizon Regularity}
The above perturbative analysis implies logarithmic growth near the classical boundary $q\to M/\lambda$. To assess the physical geometry requires evaluating the scalar curvature invariants. 
In conformal gauge, the 2-D Ricci scalar is  $R =
8e^{-2\omega}\partial_+\partial_-\omega$.  Expanding this invariant to first order in the Horndeski couplings, 
$$ R=R^{(0)}+\gamma_i R^{(1)}+\mathcal{O}(\gamma_i^2) \, , 
$$
we find that the logarithmic
divergences in the Kruskal coordinates cancel in the physical curvature (details in Appendix~\ref{app:curvature}).  The horizon curvature undergoes a finite, small shift from its classical background value of $R^{(0)}=4\lambda^2$.
Specifically, the first-order corrections evaluated at the horizon are (details in Appendix~\ref{app:curvature}):
$$R^{(1)}\big|_{\rm hor}^{G_3} = -\frac{8\lambda^7}{M^3}, \qquad
  R^{(1)}\big|_{\rm hor}^{G_5} =
  \frac{8\lambda^5}{\Lambda^2 M^3} - \frac{16\lambda^5}{M}.$$
Furthermore, the curvature-gradient invariant vanishes at the
horizon for all sectors, $(\nabla R)^2\big|_{\rm hor}=0$, precisely
mirroring the unperturbed CGHS BH.  As shown in Appendix~\ref{app:curvature}, since the ratio $\gamma_i
R^{(1)}/R^{(0)}\sim\mathcal{O}(\gamma_i)$ remains
less than $1$ throughout the near-horizon regime, the perturbative
expansion is self-consistent.  The logarithmic divergence in
the auxiliary fields is thus a pure coordinate artifact; the active
Horndeski sectors deform the near-horizon geometry but
preserve a smooth, non-singular event horizon. 

\begin{figure}[htbp]
\centering
\includegraphics[width=\columnwidth]{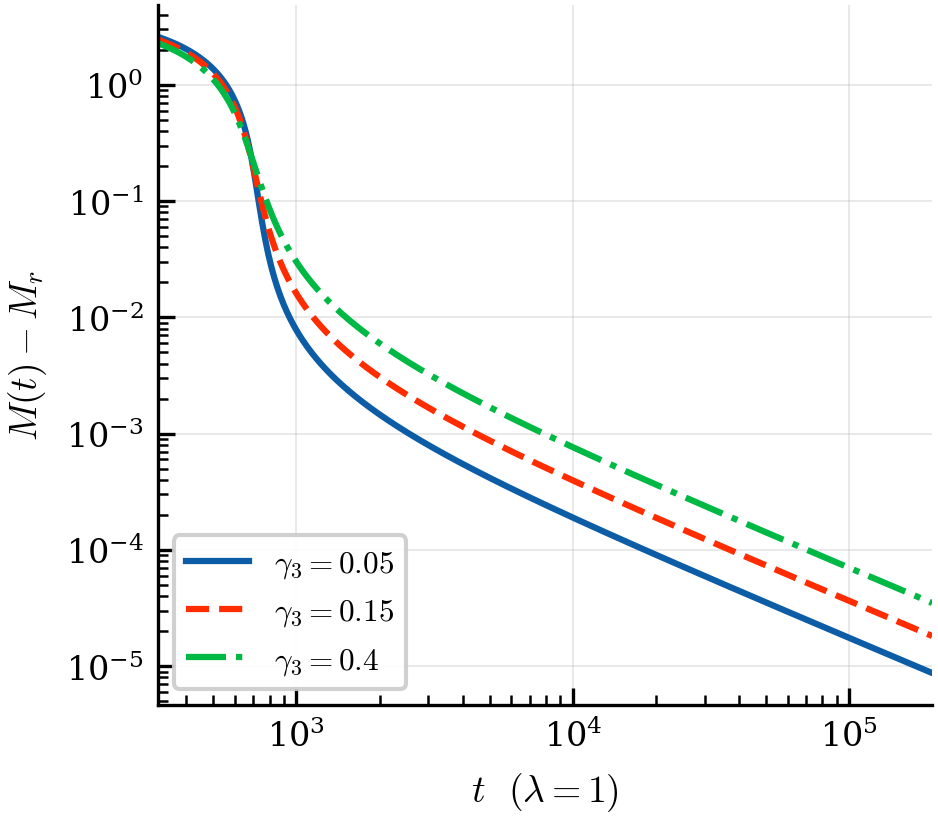}
\caption{Late-time evolution of $M(t)-M_{\rm r}$ for the $G_3$-corrected CGHS BH ($\lambda=1$, $M_0=5$). All curves parallel the analytical $t^{-1}$ slope (dashed grey), demonstrating an infinite-lifetime power-law freeze-out driven by the double-pole structure of the semiclassical luminosity.}
\label{fig:Mvst_latetime}
\end{figure}

\begin{figure}[htbp]
\centering
\includegraphics[width=\columnwidth]{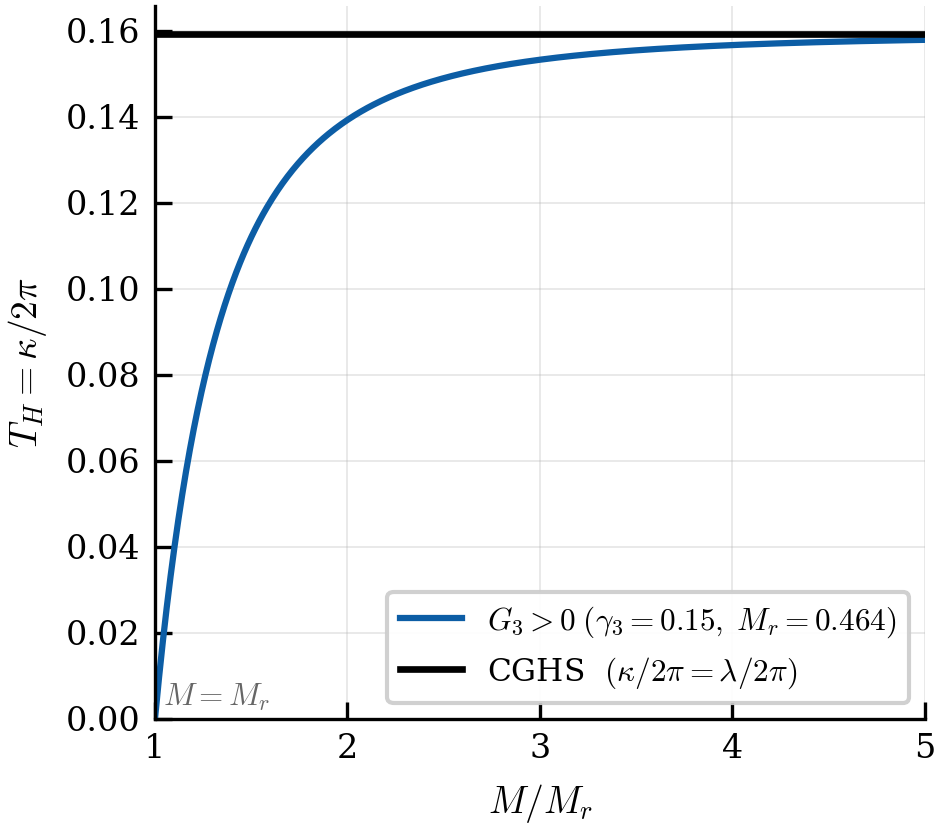}
\caption{Hawking temperature $T_H$ versus normalized mass $M/M_{\rm r}$ for the $G_3$ sector ($\gamma_3=0.15$). $T_H$ vanishes continuously at the remnant threshold ($M/M_{\rm r}=1$) and recovers the constant CGHS baseline (solid black) in the macroscopic limit $M\gg M_{\rm r}$.}
\label{fig:HawTempVsTime}
\end{figure}

\section{Temperature Shift and the Cold Remnant}
\label{sec:remnant_main}

The deformed near-horizon geometry alters the semiclassical thermodynamics. While the Kruskal profiles $A_i(q)$ and $W_i(q)$ exhibit logarithmic divergences at the horizon, this logarithm is a coordinate artifact representing the linear Taylor expansion of a modified power-law lapse function. 

To explicitly demonstrate what is resummed and how, we evaluate the near-horizon conformal factor. As detailed in Appendix~\ref{app:shift}, the perturbed geometry contains a logarithmic divergence that splits symmetrically between the null directions. This divergence is absorbed by transforming to a shifted Kruskal frame $(Y^+,Y^-)$ defined by the power-law map:
\begin{equation}
X^\pm = (Y^\pm)^{1+s_i}, \quad s_i \equiv \gamma_i\,\frac{\lambda}{M}\,w_c,
\end{equation}
where $w_c$ is the coefficient of the logarithmic divergence in $W_i(q)$. The Jacobian of this transformation precisely cancels the logarithm in the invariant dilaton, rendering the metric regular at the new horizon $Y^+Y^-=0$. 

Physically, this coordinate shift resums the perturbed lapse function. Near the horizon, the linear expansion $f \propto \epsilon [1 - s_i \ln \epsilon]$ exponentiates into a modified power law $f \propto \epsilon^{1-s_i}$. Since, the Kruskal coordinates satisfy $X^\pm \sim \pm e^{\pm \kappa r_*}$, this exponentiation corresponds directly to a fractional shift in the surface gravity, $\delta \kappa / \kappa = -s_i$. Evaluating this shift using the exact closed-form solutions (see Appendix~\ref{app:shift}) yields the corrected surface gravity. For the $G_3$ sector ($\gamma_3>0$), the Horndeski coupling decreases the Hawking temperature as BH evaporates, yielding a fractional shift $\delta T_H/T_H = -2\gamma_3\lambda^5/3M^3$.

Resumming this exact linear-order shift produces the corrected temperature profile:
\begin{equation}
T_H(M) = \frac{\lambda}{2\pi}\left(1-\frac{M_{\rm r}^3}{M^3}\right),
\label{eq:TH_resummed}
\end{equation}
where $M_{\rm r}\equiv(2\gamma_3\lambda^5/3)^{1/3}$ is the critical mass scale.  This is a key result of this work regarding which we want to discuss the following points: First, the validity of this resummation is dictated by the EFT parameter $(M_{\rm r}/M)^3$. Equation~\eqref{eq:TH_resummed} is under strict perturbative control in the regime $M \gg M_{\rm r}$, where it matches the exact linear-order calculation. As $M \to M_{\rm r}$, the expansion parameter becomes $\mathcal{O}(1)$. Therefore, Eq.~\eqref{eq:TH_resummed} functions as an extrapolation of the leading-order physics to the EFT boundary, rather than an exact non-perturbative derivation.

Second, the Stefan-Boltzmann luminosity, governed by $dM/dt=-(\pi/12)T_H^2$, determines BH lifetime. The dynamical quenching of $T_H$ forces the evaporation to halt. Expanding near the critical mass, the evaporation law transitions to an asymptotic power-law freeze-out:
\begin{equation}
M(t) - M_{\rm r} \sim \frac{1}{t}.
\end{equation}
As shown in Fig.~\ref{fig:Mvst_latetime}, BH requires infinite proper time to reach $M_{\rm r}$, asymptotically settling into a stable, macroscopic cold remnant of non-zero mass whose temperature vanishes in the limit (plotted in Fig.~\ref{fig:HawTempVsTime}). The full evaporation history and the companion luminosity curve are detailed in Appendix~\ref{app:remnant}.

Lastly, while the horizon is rendered smooth, the interior classical singularity at $q\to0$ is not resolved perturbatively. The curvature corrections diverge as $q^{-5}$, accelerating the perturbative breakdown inside BH and indicating the need for a non-perturbative ultraviolet completion at the scale $q_*\sim (\gamma_3 M\lambda)^{1/4}$. Appendix~\ref{app:shift} contains the near-singularity expansion of $R^{(1)}$ and the corresponding breakdown scale analysis.

\section{Conformal Stress Tensor and Gravitational Anomaly}
\label{sec:GravAnomaly}

To verify this kinematic evaporation from microscopic quantum
principles, we compute the exact semiclassical stress tensor.  In the conformal gauge, integrating the conservation equations
$\nabla_\mu\langle T^{\mu\nu}\rangle=0$ subject to the 2-D
Duff trace anomaly~\cite{Duff:1977ay,Christensen:1977jc},
$\langle T^\mu{}_\mu\rangle=R/24\pi$, completely determines the
energy-momentum dynamics. 
Appendix~\ref{app:anomaly} contains the detailed calculations of the two null conservation equations, their explicit Christoffel-symbol origin, and the resulting closed-form stress tensor $\langle T_{+-}\rangle$, $\langle T_{\pm\pm}\rangle$. 

\begin{figure}[htbp]
\centering
\includegraphics[width=\columnwidth]{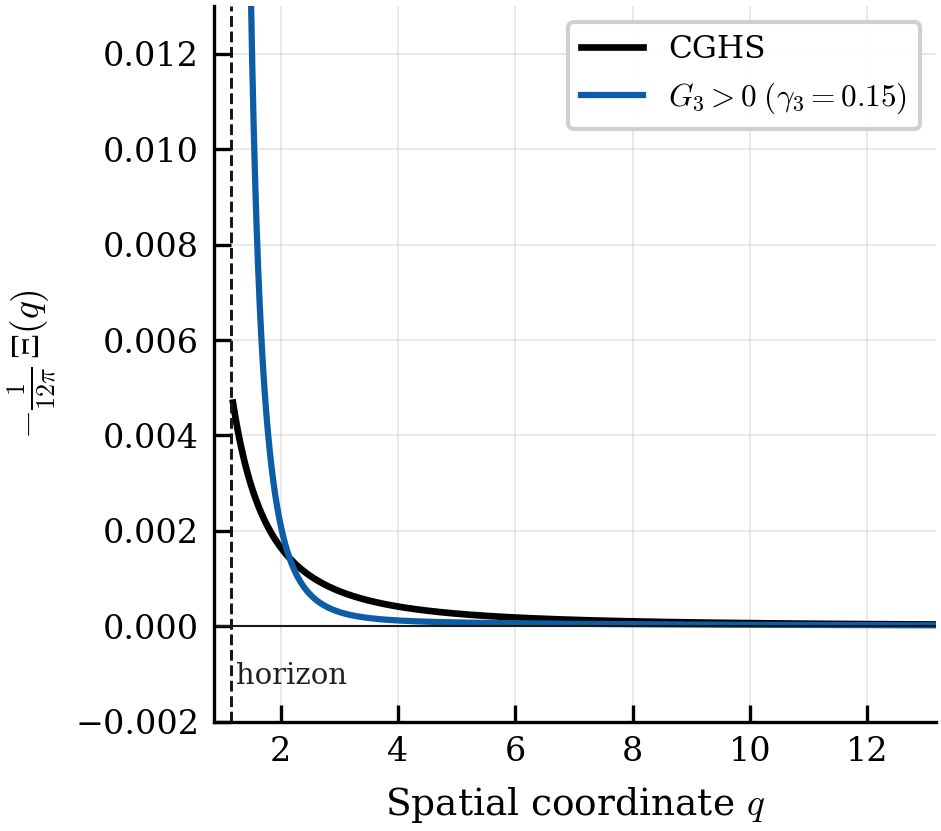}
\caption{Geometric, state-independent part $-\frac{1}{12\pi}\Xi(q)$ of the outgoing stress-tensor component, the full component being $\langle T_{--}\rangle=-\frac{1}{12\pi}(X^+)^2\Xi(q)+t_-(x^-)$, versus the spatial Kruskal coordinate $q$ ($M=2.5M_{\rm r}$, $\gamma_3=0.15$). The profile vanishes at spatial infinity, and the CGHS background curve remains finite at the horizon. The divergence of the first-order Horndeski correction as $q\to M/\lambda$, which grows as $\epsilon^{-2}$, isolates the coordinate degeneration of the Kruskal frame, a gauge artifact that cancels in physical curvature invariants.}
\label{fig:StressTensor}
\end{figure}

The outgoing asymptotic flux is carried entirely by the state-dependent integration function $t_-(x^-)$ of Eq.~\eqref{eq:app_Tpp_solved}, which in the Unruh state is fixed by the corrected surface gravity and therefore matches the macroscopic luminosity shift, confirming that the backreacted stress tensor radiates less energy as it cools. The remaining, geometric part of $\langle T_{--}\rangle$ is shown in Fig.~\ref{fig:StressTensor}. In conformal gauge, the off-diagonal component $\langle T_{+-}\rangle = -\frac{1}{96\pi} e^{2\omega} R$ is fixed algebraically by the 2-D trace anomaly. As $q \to \infty$, both this geometric part of $\langle T_{--}\rangle$ and $\langle T_{+-}\rangle$ decay rapidly to zero, explicitly confirming the preservation of asymptotic flatness in the presence of Horndeski couplings. Approaching the horizon ($q \to M/\lambda$), the growth visible in the first-order $G_3$ correction to the geometric part of $\langle T_{--}\rangle$, which diverges as $\epsilon^{-2}$ rather than logarithmically, isolates the coordinate degeneration of the baseline Kruskal frame. Because the stress-energy tensor components are coordinate-dependent field representations, this localized divergence is a non-invariant gauge artifact that leaves physical observables—such as the scalar curvature $R$—completely regular. The explicit near-horizon profiles of both stress-tensor components for the $G_3$ and $G_5$ sectors, including the companion component $\langle T_{+-}\rangle$ plotted in Fig.~\ref{fig:app_Tpm_vs_q} of Appendix~\ref{app:anomaly}, are tabulated there as well. 

To confirm this, we derived the modified flux via the Robinson-Wilczek mechanism~\cite{Robinson:2005pd}.  By constructing the invariant radial tortoise coordinate $r$ from the exact geometric identity $R=-f''(r)$, where $f$ is the Schwarzschild-type lapse function defined via $ds^2=-f(r)\,dt^2+dr^2/f(r)$, and integrating out the near-horizon ingoing modes, the effective 2-D theory becomes chiral.  General covariance is consequently broken by a purely timelike gravitational anomaly $A_t=\partial_r N^r{}_t$, where $N^r{}_t=\frac{1}{192\pi}(f'^2+f''f)$ is the anomalous flux component of the effective chiral stress tensor (defined explicitly in Appendix~\ref{app:anomaly}).  Integrating this localized anomaly exactly yields the compensating flux required to restore diffeomorphism invariance:
$$\int_{r_H}^{\infty} A_t\,dr = -\frac{\kappa^2}{48\pi}.$$
The construction of $r$ from $R=-f''(r)$, the resulting Jacobian ODE, the explicit background and first-order corrected anomaly $A_t^{(0)}+\gamma_3 A_t^{(1)}$, and the verification of the integrated-anomaly identity above are all carried out in Appendix~\ref{app:anomaly}.

Figure~\ref{fig:Anomaly} illustrates the radial profile of the total backreacted gravitational anomaly $A_t(q)$ across the valid perturbative regime ($M \ge 3M_{\rm r}$). On the unperturbed background, the anomaly $A_t^{(0)}(q)$ interpolates smoothly from its universal horizon value $A_t^{(0)}|_{r_H} = -\lambda^3 / 8\pi$ to zero at spatial infinity. When $G_3 \neq 0$, the first-order corrections $A_t^{(1)}$ systematically suppress the magnitude of $A_t(q)$ near the horizon. As the mass $M$ decreases toward the remnant threshold $M_{\rm r}$, this anomaly quenching becomes increasingly pronounced. Integrating $A_t(q)$ over the exterior spacetime $r \in [r_H, \infty)$ strictly reproduces the diminished asymptotic flux $\Phi = \kappa^2 / 48\pi$. This establishes an independent, microscopic proof that the chiral gravitational anomaly dynamically tracks the thermodynamic cooling and eventual freeze-out of the BH.

\begin{figure}[htbp]
\centering
\newtext{%
\includegraphics[width=\columnwidth]{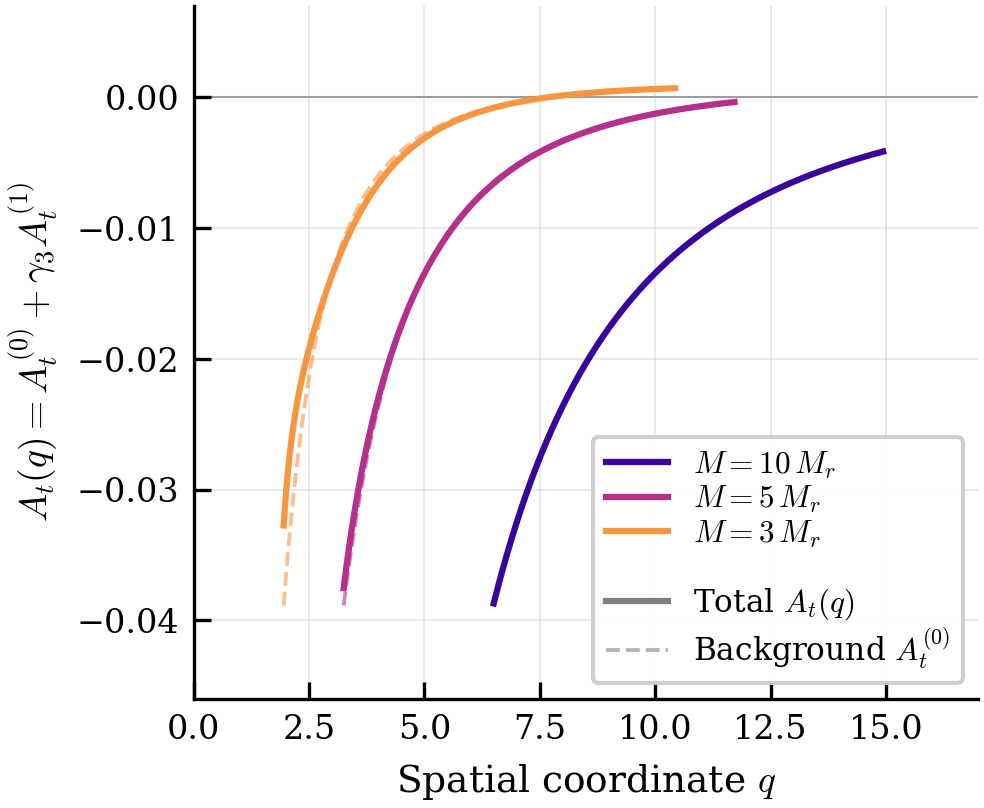}}
\caption{Total backreacted gravitational anomaly $A_t(q)$ in the perturbative regime ($M\geq 3M_{\rm r}$) for the $G_3$ sector. Horndeski corrections reduce the near-horizon anomaly relative to the CGHS background (dotted lines), reproducing the suppressed asymptotic flux $\Phi = \kappa^2/48\pi$ and confirming the cold-remnant evaporation rate.}
\label{fig:Anomaly}
\end{figure}

\newtext{%
\begin{figure*}[htbp]
\centering
\subfigure[Entropy $S(u)$]{%
  \includegraphics[width=\columnwidth]{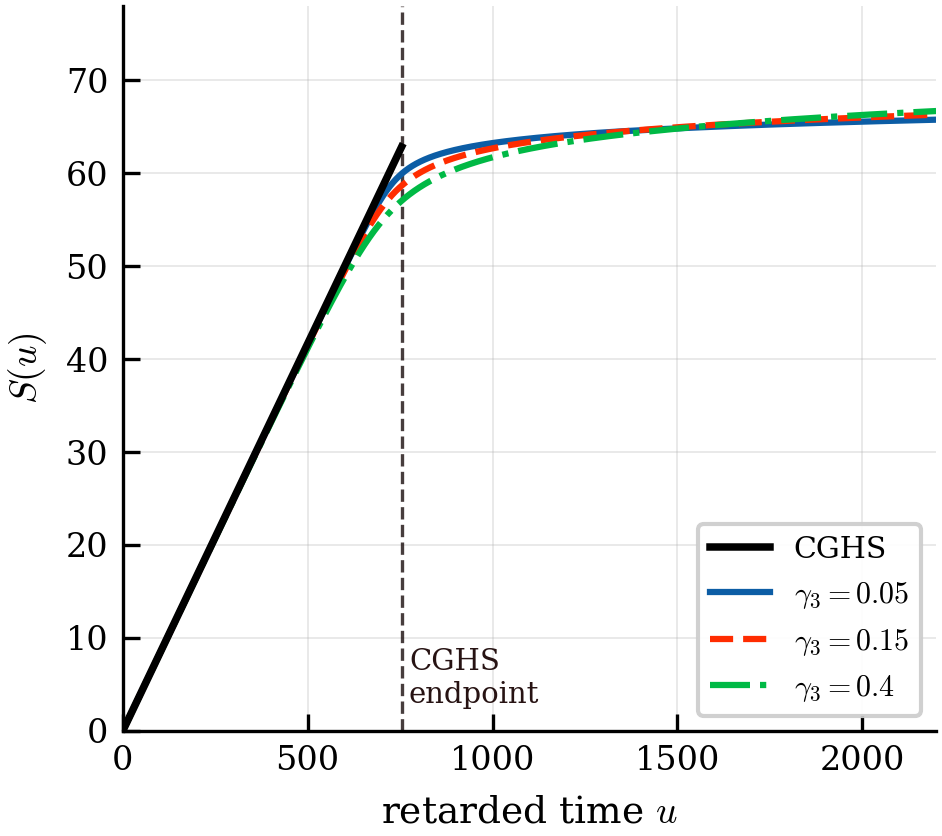}%
  \label{subfig:Szoom}}%
\hfill
\subfigure[Production rate $dS/du$]{%
  \includegraphics[width=\columnwidth]{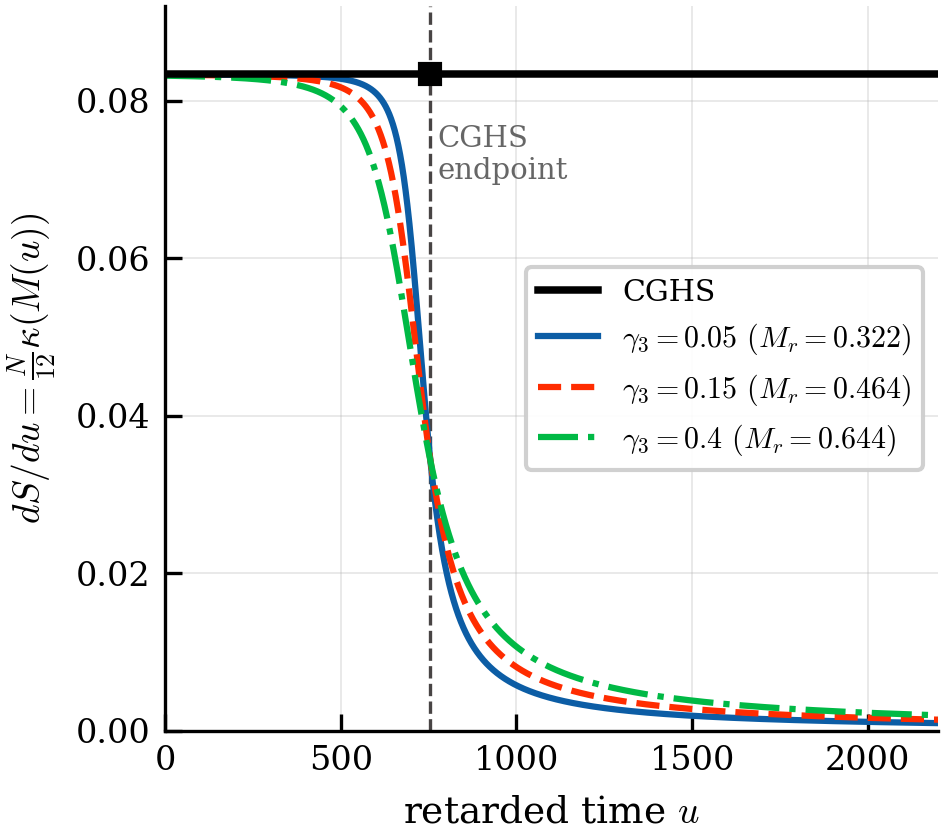}%
  \label{subfig:dSduzoom}}
\caption{%
Fine-grained entanglement entropy of the Hawking radiation
($\lambda=N=1$, $M_0=5$) in the vicinity of the CGHS evaporation
endpoint $u_f\approx754\,\lambda^{-1}$ (vertical dotted line).
(a) Entropy $S(u)$: the CGHS model (solid black) terminates abruptly
at $S_{\rm CGHS}^{\rm final}=4\pi M_0/\lambda\approx62.8$, while
each Horndeski-corrected model bends below the CGHS trajectory at
$u\lesssim u_f$ and then continues to grow logarithmically for
$u>u_f$, transitioning to $S(u)\simeq S_{\rm sat}+(4\pi NM_{\rm
r}/3\lambda)\ln u$.
(b) Entropy production rate $dS/du=(N/12)\kappa(M(u))$: the curves
peel away from the constant CGHS baseline $N\lambda/12$ at
mass-dependent rates, collapsing toward zero as $\kappa\to0$ at
the cold remnant.  The larger the coupling $\gamma_3$ (and hence the
larger $M_{\rm r}$), the earlier the rate departs from the CGHS
constant and the sooner the production switches off.}
\label{fig:entropy}
\end{figure*}}

\section{Entanglement Entropy of the Hawking Radiation}
\label{sec:Entropy}

To understand the information-theoretic implications of the cold
remnant, we evaluate the fine-grained entanglement entropy of the
Hawking radiation~\cite{Kumar:2015bha,Chandran:2020gcd,Chandran:2023ogt}. In a semiclassical evaporating background, the outgoing Hawking radiation collected at future null infinity is entangled with the modes left behind the horizon. Because the effective field theory is a $(1{+}1)$-dimensional conformal field theory (CFT), this entanglement is fixed entirely by the universal short-distance structure of the CFT vacuum together with the causal ray-tracing map between the horizon and infinity, independent of further microphysical input. Specifically, for $N$ free scalar fields, the renormalised entropy of the right-moving radiation collected by a distant observer over a retarded-time interval $[u_1,u_2]$ is governed by the Holzhey-Larsen-Wilczek (HLW)~\cite{Holzhey:1994we}/Fiola-Preskill-Strominger-Trivedi (FPST)~\cite{Fiola:1994ir} formula:
\begin{equation}
S_{\rm ren} = \frac{N}{12}\ln\!\left[\frac{\big(U(u_2)-U(u_1)\big)^2}{U'(u_1)U'(u_2)(u_2-u_1)^2}\right] \, ,
\end{equation}
where the ray-tracing relation $U(u)$ maps the Kruskal coordinate to the asymptotic retarded time $u$. For a slowly evaporating BH, the surface gravity drifts adiabatically, $\kappa=\kappa(M(u))$. The growing contribution to the entropy is dominated by the redshift of the late endpoint, $-\frac{N}{12}\ln U'(u)$. This yields a universal entropy production rate proportional to the instantaneous surface gravity~\cite{Calabrese:2004eu}:
\begin{equation}
\frac{dS}{du} = \frac{N}{12}\kappa(M).   
\end{equation}
In the unperturbed CGHS model, constant surface gravity ($\kappa=\lambda$) leads to linear entropy growth $S\propto u$. With the constant-rate law $dM/du=-\lambda^2/48\pi$, this integrates to $S = (4\pi N/\lambda)(M_0-M)$, terminating abruptly at a finite maximum entropy $S_{\rm CGHS}^{\rm final} = 4\pi NM_0/\lambda$ when the BH evaporates completely. 

In our Horndeski-corrected model, the surface gravity dynamically vanishes as $M\to M_{\rm r}$. Coupling the production rate to the modified evaporation law $dM/du = -\kappa(M)^2/48\pi$, we find the closed relation $dS/dM = -4\pi N / \kappa(M)$. As shown in Appendix~\ref{app:entropy}, integrating this reveals that $S(M)$ departs from the linear CGHS trajectory, possessing a logarithmic divergence as the mass approaches the remnant.

To determine the temporal behavior at late times, we expand the dynamics near the remnant limit, $M=M_{\rm r}+\epsilon$. The mass perturbation, governed by a Riccati-type evaporation equation, decays as $\epsilon(u)\simeq 16\pi M_{\rm r}^2/3\lambda^2 u$. This causes the surface gravity to fall as $\kappa(u)\simeq 16\pi M_{\rm r}/\lambda u$. Consequently, the entropy production rate decays as $u^{-1}$, and the entanglement entropy transitions to logarithmic growth asymptotically:
\begin{equation}
S(u) \simeq S_{\rm sat} + \frac{4\pi N M_{\rm r}}{3\lambda}\ln u.
\label{eq:Su_log_main}
\end{equation}
This linear-to-logarithmic crossover is the entanglement-entropy analogue of the length/temperature crossover familiar from the Calabrese-Cardy formula~\cite{Calabrese:2004eu} underlying the production-rate relation: just as a CFT interval's entropy crosses over from thermal, volume-law growth to logarithmic growth once the relevant spatial scale becomes small compared to the thermal correlation length, here the vanishing surface gravity $\kappa(u)\to0$ suppresses the production rate from constant to $u^{-1}$, driving an identical crossover in $S(u)$.

Unlike the standard semiclassical Page curve where entropy abruptly stops growing upon complete evaporation, the entropy in the $G_3$ sector never reaches a finite maximum. The radiation production switches off, and the entanglement entropy is permanently trapped by the infinite lifetime of the long-lived cold remnant. This behavior is illustrated in Fig.~\ref{fig:entropy}: panel (a) shows the entropy trajectory $S(u)$ near the CGHS evaporation endpoint $u_f\approx 754\,\lambda^{-1}$, and panel (b) shows the entropy production rate $dS/du$, demonstrating how each Horndeski curve peels away from the constant CGHS baseline as the BH cools toward the remnant. 

We note, however, that a genuine finite-dimensional remnant can only purify a \emph{bounded} amount of entanglement; the unbounded logarithmic growth of Eq.~\eqref{eq:Su_log_main} should therefore be understood as tracking the fine-grained entropy of the emitted radiation within the leading-order semiclassical treatment, rather than as a literal microstate count of the remnant itself. Reconciling these descriptions requires a genuinely non-perturbative accounting of the remnant's internal degrees of freedom, which lies beyond the present analysis.

It is important to compare and contrast the CGHS and Horndeski-corrected BH scenarios. The CGHS BH radiates at a high, constant rate but dies quickly at $u_f\approx750$. However, the Horndeski BH radiates its final fraction of mass over an infinite time window. Even though the temperature approaches zero, the horizon continues emitting extremely low-frequency, highly redshifted, correlated Hawking pairs for eternity. Because the asymptotic observer's Hilbert space accumulates these soft pairs forever, the integrated capacity for entanglement exceeds the short-lived CGHS burst (cf.\ Fig.~\ref{fig:entropy}(a)). 

At first sight, this seems paradoxical: the Horndeski BH radiates strictly \emph{less} total mass than the CGHS BH (it retains the finite remnant mass $M_{\rm r}$), yet it produces \emph{more} total entanglement entropy over the course of its evolution. This tension is resolved directly by evaluating the entropy produced per unit mass radiated:
$$\frac{dS}{|dM|} = \frac{4\pi N}{\kappa(M)}.$$
This is exactly the thermodynamic entropy $dS=|dM|/T_H$ released by a body radiating at temperature $T_H$. Because $\kappa(M) < \lambda$ at every mass $M > M_{\rm r}$, the efficiency exceeds the constant CGHS value $4\pi N/\lambda$ throughout the entire evaporation history. As the BH cools toward absolute zero ($M \to M_{\rm r}^+$), this entropy yield diverges. The Horndeski BH is therefore a strictly less efficient radiator of mass, but a strictly more efficient producer of entropy. An inefficient but eternal radiator can out-produce an efficient but short-lived one in total entanglement, without ever violating the strictly finite energy budget $M_0-M_{\rm r}$ available to it.

\section{Conclusions}
In this work, we established that higher-derivative Horndeski corrections to the CGHS model modify the late-time thermodynamics of Hawking evaporation. The calculation of scalar curvature invariants proves that the logarithmic divergences in the perturbed Kruskal frame are coordinate artifacts; the corrected event horizon is smooth.  Instead, these geometric deformations dictate a mass-dependent shift in the surface gravity.

For positive $G_3$ couplings, this shift induces a dynamic cooling
effect that strictly quenches the Hawking flux as the mass approaches a
critical threshold.  The exact consistency between the macroscopic
luminosity and the microscopic fluxes derived independently via the
2D conformal anomaly and the chiral Robinson-Wilczek gravitational
anomaly confirms the kinematic stability of the resulting cold
remnant.  Since the evaporation asymptotically freezes, the
entanglement entropy of the radiation transitions to logarithmic growth,
indicating that the remnant acts as a stable reservoir for the trapped
interior entanglement.

While the exterior event horizon remains smooth, our perturbative expansion reveals that the interior classical singularity ($q\to0$) diverges more rapidly than the CGHS geometry. Specifically, the perturbative regime breaks down at a finite interior radius $q_*$ (estimated in Appendix~\ref{app:shift}), necessitating a fully non-perturbative ultraviolet completion to resolve the deep interior. Nevertheless, this interior breakdown is causally isolated from the exterior thermodynamics. The formation of the cold remnant and the resulting logarithmic saturation of entanglement entropy remain robust predictions of the leading-order effective theory, offering a stable kinematic mechanism to preserve quantum information without sacrificing horizon regularity.

\vspace{0.5cm}
\noindent\emph{\underline{Acknowledgement:}} The authors are grateful to K.~Hari, N.~Jaiswal and T.~Parvez for their valuable discussions and feedback on the earlier draft. The work is supported by ANRF-Advanced Research Grant (ANRF/ARG/2025/001514/PS). This work is part of the
Undergraduate project of AS.

\appendix

\section{4-D Horndeski action}
\label{app:action}

This appendix gives the derivation of the 2-D action in Eq.~\eqref{eq:S2}, used throughout the main text. We start from the full 4-D Horndeski Lagrangian and end at the target space fixed in the CGHS gauge, quoted below Eq.~\eqref{eq:selection}.

\subsection{Spherical reduction of the Horndeski densities}

The full 4D Horndeski action~\eqref{eq:Horndeski} is assembled from four sector Lagrangian densities. With $X\equiv-\frac{1}{2}\nabla_\mu\Phi\nabla^\mu\Phi$, $G_{i,X}\equiv\partial G_i/\partial X$, and $G^{(4D)}_{\mu\nu}$ the 4D Einstein tensor, the explicit forms are:
\begin{equation}
\begin{aligned}
\mathcal{L}_2 &= G_2(\Phi,X)\,, \\
\mathcal{L}_3 &= -G_3(\Phi,X)\,\Box_4\Phi\,, \\
\mathcal{L}_4 &= G_4(\Phi,X)\,R^{(4D)} + G_{4X}\bigl[(\Box_4\Phi)^2 - (\nabla_\mu\nabla_\nu\Phi)^2\bigr]\,, \\
\mathcal{L}_5 &= G_5(\Phi,X)\,G^{(4D)}_{\mu\nu}\nabla^\mu\nabla^\nu\Phi \\
&\quad -\frac{1}{6}G_{5X}\bigl[(\Box_4 \Phi)^3  - 3(\Box_4 \Phi)(\nabla_\mu\nabla_\nu\Phi)^2 + 2(\nabla\nabla\Phi)^3\bigr]\,,
\end{aligned}
\label{eq:L_explicit}
\end{equation}
where $(\nabla\nabla\Phi)^3\equiv \nabla_\mu^{\ \nu}\Phi\,\nabla_\nu^{\ \rho}\Phi\,\nabla_\rho^{\ \mu}\Phi$ and $\Box_4$ corresponds to d'Alembertian for 4-D space-time.

After spherical reduction~\eqref{eq:ansatz}, integrating out the $S^2$ angular directions and absorbing the resulting $4\pi$ volume factor into the coupling definitions, each sector projects to a 2-D density $\mathcal{L}_{2\rm D}=\sum_i L_{G_i}$ with:
\begin{equation}
\begin{aligned}
L_{G_2} &= 4\pi\rho^2\,G_2\,, \\
L_{G_3} &= -4\pi G_3\bigl[\rho^2\Box\Phi+2\rho Z\bigr]\,, \\
L_{G_4} &= 4\pi G_4\bigl[\rho^2 R-4\rho\Box\rho+2(1-Y)\bigr]\,, \\
L_{G_{4X}} &= 4\pi G_{4X}\bigl[\rho^2\{(\Box\Phi)^2-\mathcal{ I}\} + 4\rho Z\,\Box\Phi+2Z^2\bigr]\,, \\
L_{G_5} &= 4\pi G_5\bigl[-2\rho(W-\Box\rho\,\Box\Phi) - (1-Y)\Box\Phi \\ &\quad + 2Z\Box\rho-\rho Z R\bigr]\,, \\
L_{G_{5X}} &= -\frac{2\pi}{3}G_{5X}\bigl[\rho^2\{(\Box\Phi)^3-3\Box\Phi\,\mathcal{ I}+2\mathcal{C}\} \\
&\quad + 6\rho Z\{(\Box\Phi)^2-\mathcal{ I}\} + 6Z^2\Box\Phi\bigr]\,,
\end{aligned}
\label{eq:LG_all}
\end{equation}
with the 2D kinematic bilinears defined in \eqref{eq:kin_1}--\eqref{eq:kin_2}. The ghost-freedom selection rule $G_{4,X}=G_{5,X}=0$ eliminates $L_{G_{4X}}$ and $L_{G_{5X}}$ from the surviving action~\eqref{eq:S2}, leaving $\mathcal{L}_{2\rm D}=L_{G_2}+L_{G_3}+L_{G_4}+L_{G_5}$ with $G_2(\Phi,X),G_3(\Phi,X)$ unrestricted and $G_4(\Phi),G_5(\Phi)$ purely field-dependent.

\subsection{Brans-Dicke gauge and recovery of the CGHS target space}
\label{app:cghs_gauge}

We now fix the residual freedom in $\mathcal{L}_{2\rm D}$ by adopting the parametrization (see \citep{Mandal:2023kpu})
\begin{equation}
\Phi = e^{2\varphi}, \qquad \rho = \Lambda\,e^{-2\varphi}, \qquad G_4(\Phi)=\Phi,
\label{eq:app_BD_gauge}
\end{equation}
where $\Lambda$ is a constant of dimension length fixing the overall normalization of the areal radius and $\varphi$ is the CGHS dilaton. This choice is not arbitrary but is precisely the parametrization under which the unperturbed ($G_3=G_5=0$) theory reduces to the linear-dilaton vacuum of the classical CGHS model, and it is the gauge in which the Kruskal fields $u=e^{-2\varphi}$, $v=e^{-2\omega}$ used throughout Appendix~\ref{app:eom} are naturally defined.

Because $\rho=\Lambda e^{-2\varphi}$ is itself built from $\varphi$, every kinematic bilinear appearing in Eq.~\eqref{eq:LG_all} inherits a definite dependence on $\varphi$-derivatives. Writing $X_\varphi\equiv-\tfrac12\nabla^a\varphi\nabla_a\varphi$ and $\Box\varphi\equiv\nabla^a\nabla_a\varphi$, direct differentiation of $\Phi=e^{2\varphi}$ and $\rho=\Lambda e^{-2\varphi}$ gives the induced relations
\begin{equation}
\begin{aligned}
X_\Phi &= 4e^{4\varphi}X_\varphi, \quad Z = 8\Lambda X_\varphi, \quad Y=-8\rho^2 X_\varphi, \\
\Box\Phi &= 2\Phi\big(\Box\varphi-4X_\varphi\big), \quad
\Box\rho = -2\rho\big(\Box\varphi+4X_\varphi\big),
\end{aligned}
\label{eq:app_induced_relations}
\end{equation}
where $X_\Phi\equiv-\tfrac12\nabla^a\Phi\nabla_a\Phi$ denotes $X$ evaluated on $\Phi$. These five relations needed to rewrite every term of Eq.~\eqref{eq:LG_all} entirely in terms of $\varphi$. Also, take $G_2(\Phi,X)=\Omega(\Phi)X_\Phi-V(\Phi)$, a general non-canonical kinetic term. Substituting $\rho^2=\Lambda^2e^{-4\varphi}$ directly into $L_{G_2}=4\pi\rho^2 G_2$ of Eq.~\eqref{eq:LG_all} gives,
\begin{align}
L_{G_2} &= 4\pi\Lambda^2 e^{-4\varphi}\Big[\Omega(\Phi)\,X_\Phi - V(\Phi)\Big] \nonumber \\
&= 4\pi\Lambda^2 e^{-4\varphi}\Omega(\Phi)X_\Phi - 4\pi\Lambda^2 e^{-4\varphi}V(\Phi).
\label{eq:app_LG2_step1}
\end{align}
Now substitute $X_\Phi=4e^{4\varphi}X_\varphi$ from Eq.~\eqref{eq:app_induced_relations} into the first term only:
\begin{equation}
4\pi\Lambda^2 e^{-4\varphi}\,\Omega(\Phi)\cdot 4e^{4\varphi}X_\varphi
= 16\pi\Lambda^2\,\Omega(\Phi)\,X_\varphi .
\end{equation}
Collecting both pieces and stripping the overall $4\pi$ ,
\begin{equation}
\hat L_{G_2} \equiv \frac{L_{G_2}}{4\pi} = \Lambda^2\Big[4\,\Omega(\Phi)\,X_\varphi - e^{-4\varphi}V(\Phi)\Big].
\label{eq:app_LG2_final}
\end{equation}
Similarly, insert Eq.~\eqref{eq:app_induced_relations} into Eq.~\eqref{eq:LG_all} and collect powers of $e^{\pm2\varphi}$ gives, for the $G_3$ sector with $G_3=G_3(\Phi,X_\Phi)$,
\begin{equation}
\hat L_{G_3} = -2\Lambda^2 e^{-2\varphi}\,G_3(\Phi,X_\Phi)\,\big(\Box\varphi+4X_\varphi\big).
\label{eq:app_LG3_final}
\end{equation}
For $G_4=\Phi$, direct substitution of $\rho^2 R=\Lambda^2e^{-4\varphi}R\cdot e^{2\varphi}\cdot e^{-2\varphi}$ together with $\Box\rho$ from Eq.~\eqref{eq:app_induced_relations} gives
\begin{equation}
\hat L_{G_4} = \Lambda^2 e^{-2\varphi}R + 8\Lambda^2 e^{-2\varphi}\Box\varphi + 48\Lambda^2 e^{-2\varphi}X_\varphi + 2e^{2\varphi},
\label{eq:app_LG4_final}
\end{equation}
and for $G_5=G_5(\Phi)$, substituting $Y,Z,\Box\rho,\Box\Phi$ from Eq.~\eqref{eq:app_induced_relations} into $L_{G_5}=4\pi G_5[-2\rho(W-\Box\rho\Box\Phi)-(1-Y)\Box\Phi+2Z\Box\rho-\rho Z \, R]$, and writing $\mathcal{ I}^{(\varphi)}\equiv\nabla_a\nabla_b\varphi\nabla^a\nabla^b\varphi$ for the dilaton Hessian trace, gives
\begin{equation}
\begin{aligned}
\hat L_{G_5} = G_5(\Phi)\Big\{&-8\Lambda^2 e^{-2\varphi}\big[(\Box\varphi)^2-\mathcal{ I}^{(\varphi)}+6X_\varphi\Box\varphi \\
&+8X_\varphi^2+X_\varphi R\big] -2e^{2\varphi}\big(\Box\varphi-4X_\varphi\big)\Big\}.
\end{aligned}
\label{eq:app_LG5_final}
\end{equation}
Equations~\eqref{eq:app_LG2_final}--\eqref{eq:app_LG5_final} are the complete gauge-fixed sector Lagrangians (with the overall $4\pi$ dropped throughout, as in the main text).

The classical CGHS theory is recovered entirely from the $G_2+G_4$ piece, with the free functions in $G_2=\Omega(\Phi)X_\Phi-V(\Phi)$ fixed to
\begin{equation}
\Omega(\Phi) = -\frac{6}{\Phi}, \qquad V(\Phi) = \frac{2\Phi^3}{\Lambda^2}-4\lambda^2\Phi,
\label{eq:app_Omega_V}
\end{equation}
where $\lambda$ is the CGHS cosmological constant setting the linear-dilaton slope (see \citep{Mandal:2023kpu}). Inserting Eq.~\eqref{eq:app_Omega_V} into Eq.~\eqref{eq:app_LG2_final} and adding Eq.~\eqref{eq:app_LG4_final}, leaves exactly
\begin{equation}
\hat L_{G_2}+\hat L_{G_4} = \Lambda^2 e^{-2\varphi}\Big[R + 4(\nabla\varphi)^2 + 4\lambda^2\Big] \equiv \hat L_{\rm CGHS},
\label{eq:app_LCGHS}
\end{equation}
which is exactly the classical CGHS Lagrangian quoted below Eq.~\eqref{eq:selection} of the main text. The surviving Horndeski corrections are then carried entirely by $\hat L_{G_3}$ and $\hat L_{G_5}$ of Eqs.~\eqref{eq:app_LG3_final} and~\eqref{eq:app_LG5_final}, evaluated on the specific choices $G_3=\gamma_3\Phi X_\Phi$ and $G_5=\gamma_5\Phi$ used throughout this paper.

\section{Equations of motion}
\label{app:eom}

This appendix derives the field equations used in Sec.~III of the main text; first in the original conformal-gauge fields $(\varphi,\omega)$ in which the action~\eqref{eq:S2} is varied (Appendix~\ref{app:eom_phiomega}), then in the Kruskal-type fields $(u,v)=(e^{-2\varphi},e^{-2\omega})$ already quoted in the main text (Appendix~\ref{app:eom_uv}), then on the static slice that defines the areal coordinate $q$ (Appendix~\ref{app:static_q}), and finally linearized and solved in closed form (Appendices~\ref{app:linearized} and~\ref{app:closed_form}).

\subsection{General field equations in conformal null gauge}
\label{app:eom_phiomega}

Varying the gauge-fixed action of Appendix~\ref{app:action} with respect to $\omega$, $\varphi$, and the residual conformal constraints $g_{++},g_{--}$ produces three independent equations, denoted $\mathcal E_\omega=0$, $\mathcal E_\varphi=0$, and $\mathcal E_{++}=0$ (with $\mathcal E_{--}=0$ following from $+\leftrightarrow-$). We record them here for general $G_3(\Phi,X_\Phi)$ and $G_5(\Phi)$, using the shorthand $\varphi_\pm\equiv\partial_\pm\varphi$, $\varphi_{\pm\pm}\equiv\partial_\pm^2\varphi$, $\varphi_{+-}\equiv\partial_+\partial_-\varphi$, and likewise for $\omega$; $G_{3,\Phi}\equiv\partial G_3/\partial\Phi$, $G_{3,X}\equiv\partial G_3/\partial X_\Phi$, $G_5'\equiv dG_5/d\Phi$. Terms proportional to $G_{3,X}$ and $G_{3,XX}$ track exactly those integration-by-parts remainders that the identity below Eq.~\eqref{eq:kin_2} was designed to control; they vanish identically once $G_3$ is taken to be at most linear in $X_\Phi$, which is precisely the case $G_3=\gamma_3\Phi X_\Phi$ used throughout this paper (for which $G_{3,XX}=0$ and $G_{3,X}=\gamma_3\Phi$).

The chiral constraint reads
\begin{align}
\mathcal E_{++} &= 4\Lambda^2 e^{-2\varphi}\big(\varphi_{++}-2\omega_+\varphi_+\big) 
+ 8\Lambda^2 e^{-2\varphi}\,G_{3,\Phi}\,\varphi_+^2 \nonumber\\
& + 32\Lambda^2 e^{-2\varphi}\,G_{3,X}\,\varphi_+\varphi_-\big(6\varphi_+^2-2\omega_+\varphi_++\varphi_{++}\big)e^{-2\omega+2\varphi} \nonumber\\
& + 8\,G_5'\,\varphi_+\big(32\Lambda^2\varphi_+^2\varphi_- - 48\Lambda^2\varphi_+\varphi_-\omega_+ \nonumber\\
& + 24\Lambda^2\varphi_{++}\varphi_- + \varphi_+e^{4\varphi+2\omega}\big)e^{-4\varphi-2\omega} \nonumber\\
& + 128\Lambda^2\,G_5''\,\varphi_+^3\varphi_-\,e^{-2\omega}\ =\ 0,
\label{eq:app_Epp_phiomega}
\end{align}
the conformal constraint reads
\begin{align}
\mathcal E_\omega &= 4\Lambda^2 e^{-2\varphi}\big(\lambda^2 e^{2\omega}+4\varphi_+\varphi_--2\varphi_{+-}\big) \nonumber\\
&\quad + 64\Lambda^2\,G_{3,X}\,\varphi_+\varphi_-\big(2\varphi_+\varphi_--\varphi_{+-}\big)\,e^{2\varphi-2\omega} \nonumber\\
&\quad + 128\Lambda^2\,G_5'\,\varphi_+\varphi_-\big(2\varphi_+\varphi_--3\varphi_{+-}\big)\,e^{-2\omega} \nonumber\\
&\quad - 256\Lambda^2\,G_5''\,\varphi_+^2\varphi_-^2\,e^{2\varphi-2\omega}\ =\ 0,
\label{eq:app_Eomega_phiomega}
\end{align}
and the dilaton equation splits naturally into a CGHS piece plus additive $G_3$ and $G_5$ sector contributions, $\mathcal E_\varphi=\mathcal E_\varphi^{\rm CGHS}+\mathcal E_\varphi^{(3)}+\mathcal E_\varphi^{(5)}$, with
\begin{equation}
\mathcal E_\varphi^{\rm CGHS} = -4\Lambda^2 e^{-2\varphi}\big(\lambda^2 e^{2\omega}+4\varphi_+\varphi_--4\varphi_{+-}+2\omega_{+-}\big).
\label{eq:app_EphiCGHS_phiomega}
\end{equation}
The $G_3$-sector piece contains five distinct terms, one for each independent derivative of $G_3(\Phi,X_\Phi)$:
\begin{align}
\mathcal E_\varphi^{(3)}\Big/G_{3,\Phi} &= 16\Lambda^2\varphi_{+-},
\label{eq:app_Ephi3_a}\\[4pt]
\mathcal E_\varphi^{(3)}\Big/G_{3,\Phi\Phi} &= 16\Lambda^2 e^{2\varphi}\varphi_+\varphi_-, \label{eq:app_Ephi3_b}\\[4pt]
\mathcal E_\varphi^{(3)}\Big/G_{3,X} &= 32\Lambda^2 e^{2\varphi-2\omega}\Big(16\varphi_+^2\varphi_-^2 - 12\varphi_+^2\varphi_-\omega_- \nonumber\\
& + 6\varphi_+^2\varphi_{--} + 20\varphi_+\varphi_{+-}\varphi_- - 12\varphi_+\varphi_-^2\omega_+ \nonumber\\
& + 4\varphi_+\varphi_-\omega_+\omega_- - 2\varphi_+\varphi_-\omega_{+-} - 2\varphi_+\varphi_{--}\omega_+ \nonumber\\
& + 6\varphi_{++}\varphi_-^2 - 2\varphi_{++}\varphi_-\omega_- + \varphi_{++}\varphi_{--} - \varphi_{+-}^2\Big), \label{eq:app_Ephi3_c}\\[4pt]
\mathcal E_\varphi^{(3)}\Big/G_{3,\Phi X} &= 64\Lambda^2 e^{4\varphi-2\omega}\Big(12\varphi_+^2\varphi_-^2 - 2\varphi_+^2\varphi_-\omega_- \nonumber\\
& + \varphi_+^2\varphi_{--} - 2\varphi_+\varphi_-^2\omega_+ + \varphi_{++}\varphi_-^2\Big), \label{eq:app_Ephi3_d}\\[4pt]
\mathcal E_\varphi^{(3)}\Big/G_{3,XX} &= 256\Lambda^2 e^{6\varphi-4\omega}\varphi_+\varphi_-\Big(32\varphi_+^2\varphi_-^2 - 12\varphi_+^2\varphi_-\omega_- \nonumber\\
& + 6\varphi_+^2\varphi_{--} + 4\varphi_+\varphi_{+-}\varphi_- - 12\varphi_+\varphi_-^2\omega_+ \nonumber\\
& + 4\varphi_+\varphi_-\omega_+\omega_- - 2\varphi_+\varphi_{--}\omega_+ + 6\varphi_{++}\varphi_-^2 \nonumber\\
& - 2\varphi_{++}\varphi_-\omega_- + \varphi_{++}\varphi_{--} - \varphi_{+-}^2\Big), \label{eq:app_Ephi3_e}
\end{align}
understood as $\mathcal E_\varphi^{(3)}=G_{3,\Phi}\cdot\eqref{eq:app_Ephi3_a} + G_{3,\Phi\Phi}\cdot\eqref{eq:app_Ephi3_b} + G_{3,X}\cdot\eqref{eq:app_Ephi3_c} + G_{3,\Phi X}\cdot\eqref{eq:app_Ephi3_d} + G_{3,XX}\cdot\eqref{eq:app_Ephi3_e}$. Likewise, the $G_5$-sector piece organizes into two terms,
\begin{align}
\mathcal E_\varphi^{(5)}\Big/G_5' &= -16e^{-2\omega}\Big(32\Lambda^2\varphi_+^2\varphi_-\omega_- - 16\Lambda^2\varphi_+^2\varphi_{--} \nonumber\\
& - 64\Lambda^2\varphi_+\varphi_{+-}\varphi_- + 32\Lambda^2\varphi_+\varphi_-^2\omega_+ \nonumber\\
& - 48\Lambda^2\varphi_+\varphi_-\omega_+\omega_- + 24\Lambda^2\varphi_+\varphi_-\omega_{+-} \nonumber\\
& + 24\Lambda^2\varphi_+\varphi_{--}\omega_+ - 16\Lambda^2\varphi_{++}\varphi_-^2 \nonumber\\
& + 24\Lambda^2\varphi_{++}\varphi_-\omega_- - 12\Lambda^2\varphi_{++}\varphi_{--} \nonumber\\
& + 12\Lambda^2\varphi_{+-}^2 - 2\varphi_+\varphi_-e^{4\varphi+2\omega} - \varphi_{+-}e^{4\varphi+2\omega}\Big), \label{eq:app_Ephi5_a}\\[4pt]
\mathcal E_\varphi^{(5)}\Big/G_5'' &= 16e^{2\varphi-2\omega}\Big(48\Lambda^2\varphi_+^2\varphi_-^2 - 16\Lambda^2\varphi_+^2\varphi_-\omega_- \nonumber\\
& + 8\Lambda^2\varphi_+^2\varphi_{--} - 16\Lambda^2\varphi_+\varphi_{+-}\varphi_- - 16\Lambda^2\varphi_+\varphi_-^2\omega_+ \nonumber\\
& + 8\Lambda^2\varphi_{++}\varphi_-^2 + \varphi_+\varphi_-e^{4\varphi+2\omega}\Big), \label{eq:app_Ephi5_b}
\end{align}
with $\mathcal E_\varphi^{(5)}=G_5'\cdot\eqref{eq:app_Ephi5_a}+G_5''\cdot\eqref{eq:app_Ephi5_b}$.

Finally, since the reduction of Appendix~\ref{app:action} was performed directly on $\Phi$-dependent couplings while Eqs.~\eqref{eq:app_Epp_phiomega}--\eqref{eq:app_Ephi5_b} are written for $\varphi$-dependent ones, the two are related by the chain rule through $\Phi=e^{2\varphi}$. Writing $\widetilde G_3(\varphi,X_\varphi)\equiv G_3(\Phi,X_\Phi)$ and $\widetilde G_5(\varphi)\equiv G_5(\Phi)$, one finds that
\begin{equation}
\begin{aligned}
\widetilde G_{3,X_\varphi} &= 4\Phi^2 G_{3,X}, \\
\widetilde G_{3,\varphi} &= 2\Phi G_{3,\Phi}+16\Phi^2 X_\varphi G_{3,X}, \\
\widetilde G_5'(\varphi) &= 2\Phi G_5'(\Phi), \\
\widetilde G_5''(\varphi) &= 4\Phi\big(\Phi G_5''+G_5'\big),
\end{aligned}
\label{eq:app_dictionary}
\end{equation}
with $\Phi=e^{2\varphi}$ throughout. For the specific choices $G_3=\gamma_3\Phi X_\Phi$ and $G_5=\gamma_5\Phi$ used in this paper, Eq.~\eqref{eq:app_dictionary} gives simply $G_{3,\Phi}=\gamma_3X_\Phi$, $G_{3,X}=\gamma_3\Phi$, $G_{3,\Phi\Phi}=G_{3,\Phi X}=G_{3,XX}=0$, $G_5'=\gamma_5$, $G_5''=0$, which is exactly the simplification exploited in the Kruskal-field equations of the next subsection.

\subsection{Exact equations in Kruskal-type fields}
\label{app:eom_uv}
Substituting the specific choices $G_3=\gamma_3\Phi X_\Phi$, $G_5=\gamma_5\Phi$ into Eqs.~\eqref{eq:app_Epp_phiomega}--\eqref{eq:app_Ephi5_b}, and changing variables from $(\varphi,\omega)$ to the Kruskal-type fields $u\equiv e^{-2\varphi}$, $v\equiv e^{-2\omega}$ via the chain rule $\varphi_\pm=-u_\pm/2u$, $\omega_\pm=-v_\pm/2v$ (and similarly for the second derivatives), yields the exact equations. The subscript notation is $u_\pm\equiv\partial_\pm u$, $u_{\pm\pm}\equiv\partial_\pm^2 u$, $u_{+-}\equiv\partial_+\partial_-u$, and likewise for $v$. The conformal constraint reads
\begin{equation}
\begin{split}
\mathcal{E}_\omega &= 4\Lambda^2\frac{\lambda^2 u+v\,u_{+-}}{v} +\frac{8\Lambda^2 u_+u_-v}{u^5} \bigl[(6u^2\gamma_5+\gamma_3)\,u_{+-} \\ &-4u\gamma_5\,u_+u_-\bigr] = 0\,,
\end{split}
\label{eq:Eomega_exact}
\end{equation}
the chiral constraint reads
\begin{equation}
\begin{split}
\mathcal{E}_{++} &= \frac{2\Lambda^2}{u}\bigl(u\,u_+v_+ + u\,u_{++}v - u_+^2v\bigr) \\
&\quad + \frac{2u_+v}{u^6} \Bigl\{ \gamma_5\bigl[12\Lambda^2u^3(u_+u_-v_+ + u_{++}u_-v) \\
&\qquad - 20\Lambda^2u^2u_+^2u_-v - u^2u_+\bigr] \\
&\quad + \gamma_3\bigl[2\Lambda^2u(u_+u_-v_+ + u_{++}u_-v) - 10\Lambda^2u_+^2u_-v\bigr] \Bigr\} = 0\,, 
\end{split}
\label{eq:Epp_exact}
\end{equation}

and the dilaton equation reads $\mathcal{E}_\varphi = \mathcal{E}_\varphi^{\rm CGHS} + \gamma_3\,\mathcal{E}_\varphi^{(3)} + \gamma_5\,\mathcal{E}_\varphi^{(5)} = 0$.

The CGHS baseline is
\begin{equation}
\begin{split}
\mathcal{E}_\varphi^{\rm CGHS} &= \frac{4\Lambda^2}{uv^2}\bigl(u^2v\,v_{+-} - \lambda^2u^2v - u^2v_+v_- \\
&\quad - 2uv^2u_{+-} + v^2u_+u_-\bigr)\,.
\end{split}
\label{eq:EphiCGHS}
\end{equation}

The exact $G_3$ correction is
\begin{equation}
\begin{split}
\mathcal{E}_\varphi^{(3)} &= \frac{8\Lambda^2}{u^6}\Bigl[ u^2\bigl(u_+u_-v_{+-} + u_+u_{--}v_+ + u_{++}u_-v_- \\
&\quad + u_{++}u_{--}v - u_{+-}^2v\bigr) - 5u\bigl(u_+^2u_-v_- + u_+^2u_{--}v \\
&\quad + 2u_+u_{+-}u_-v + u_+u_-^2v_+ + u_{++}u_-^2v\bigr) \\
&\quad + 30u_+^2u_-^2v\Bigr]\,,
\end{split}
\label{eq:Ephi3}
\end{equation}

and the exact $G_5$ correction is
\begin{equation}
\begin{split}
\mathcal{E}_\varphi^{(5)} &= \frac{8}{u^6}\Bigl[ 6\Lambda^2u^4\bigl(u_+u_-v_{+-} + u_+u_{--}v_+ + u_{++}u_-v_- \\
&\quad + u_{++}u_{--}v - u_{+-}^2v\bigr) - 2\Lambda^2u^3\bigl(5u_+^2u_-v_- \\
&\quad + 5u_+^2u_{--}v + 2u_+u_{+-}u_-v + 5u_+u_-^2v_+ \\
&\quad + 5u_{++}u_-^2v\bigr) + 24\Lambda^2u^2u_+^2u_-^2v \\
&\quad - u^3u_{+-} + 2u^2u_+u_-\Bigr]\,.
\end{split}
\label{eq:Ephi5}
\end{equation}

On the Kruskal static slice $v=u$, the CGHS dilaton equation~\eqref{eq:EphiCGHS} reduces to $-4\Lambda^2(u_{+-}+\lambda^2)=0$, which is consistent with $\mathcal{E}_\omega=0$ evaluated on the background.

\subsection{Static reduction and the areal coordinate \texorpdfstring{$q$}{q}}
\label{app:static_q}
To solve Eqs.~\eqref{eq:Eomega_exact}--\eqref{eq:Ephi5} perturbatively we specialize to the static, spherically-collapsed background relevant for an eternal (or adiabatically evaporating) BH. Introduce static Kruskal coordinates $X^\pm$, in which any static field depends only on the product $\chi\equiv X^+X^-$; on such a slice the null derivatives collapse to ordinary derivatives in $\chi$,
\begin{align}
\partial_+\partial_- F = F' + \chi F'', \quad u_+u_- = \chi\,(u')^2, \nonumber \\
u_{++}u_{--}=\chi^2(u'')^2,
\label{eq:app_kruskal_static}
\end{align}
with ${}'\equiv d/d\chi$. It is convenient to trade $\chi$ for the areal-type radial variable
\begin{equation}
q \equiv \frac{M}{\lambda}-\lambda^2\chi, \qquad \frac{d}{d\chi}=-\lambda^2\frac{d}{dq},
\label{eq:app_qdef}
\end{equation}
in which the unperturbed CGHS BH solution takes the simple form
\begin{equation}
u_0 = v_0 = q, \qquad \alpha_0\equiv\frac{u_0}{v_0}=1.
\label{eq:app_cghs_q}
\end{equation}
This is the unique static slice coordinate in which the unperturbed dilaton and conformal factor coincide identically, so that a nonzero $\alpha-1$ at any order in $\gamma_i$ is an unambiguous, gauge-invariant signature of the Horndeski backreaction. The coordinate ranges over $q\to\infty$ at spatial infinity and $q\to M/\lambda$ at the classical event horizon; one directly checks, by substituting Eq.~\eqref{eq:app_cghs_q} into Eqs.~\eqref{eq:Eomega_exact} and~\eqref{eq:EphiCGHS} at $\gamma_i=0$, that both are satisfied identically.

\subsection{Linearized equations and the master equation for $A(q)$}
\label{app:linearized}

We now perturb around the background~\eqref{eq:app_cghs_q}. Write,
\begin{align}
\alpha(q) = 1+\gamma_i A(q)+O(\gamma_i^2), \nonumber \\ v(q) = q+\gamma_i W(q)+O(\gamma_i^2),
\label{eq:app_pert_ansatz}
\end{align}
with $\gamma_i\in\{\gamma_3,\gamma_5\}$, so that $u=\alpha v=q+\gamma_i[A(q)q+W(q)]+O(\gamma_i^2)$. At $O(\gamma_i^0)$ Eqs.~\eqref{eq:Eomega_exact} and~\eqref{eq:EphiCGHS} are satisfied identically by Eq.~\eqref{eq:app_cghs_q}. At $O(\gamma_i)$, substituting Eq.~\eqref{eq:app_pert_ansatz} into Eqs.~\eqref{eq:Eomega_exact} and \eqref{eq:Ephi3}/\eqref{eq:Ephi5}, using Eq.~\eqref{eq:app_kruskal_static}--\eqref{eq:app_qdef} to convert every null derivative to a $q$-derivative, and clearing an overall factor of $\Lambda^2\lambda$, gives two coupled linear ODEs for $A(q)$ and $W(q)$ in each sector. In writing them below we have further multiplied through by an overall power of $q$, namely $q^4$ in both linearized $\mathcal E_\omega$ equations and in the linearized $\mathcal E_\varphi^{(5)}$ equation, but $q^5$ in the linearized $\mathcal E_\varphi^{(3)}$ equation; this deliberate asymmetry in the normalization is the reason the decoupling combination quoted below differs between the two sectors. For $G_5=\gamma_5\Phi$,

\begin{equation}
\begin{split}
&-4q^5(\lambda q-M)A'' - 4q^4(3\lambda q-2M)A' \\
&\quad - 4q^4(\lambda q-M)W'' - 4\lambda q^4 W' \\
&\quad + 16\lambda q(\lambda q-M)(\lambda q+2M) = 0,
\end{split}
\label{eq:app_Ew5}
\end{equation}
\begin{equation}
\begin{split}
&8q^5(\lambda q-M)A'' + 8q^4(2\lambda q-M)A' \\
&\quad + 4q^4(\lambda q-M)W'' + 4\lambda q^4 W' \\
&\quad - \frac{8}{\Lambda^2}\Big(10\Lambda^2 M\lambda^2 q^2 - 4\Lambda^2 M^2\lambda q + \lambda q - 2M\Big) = 0,
\end{split}
\label{eq:app_Ep5}
\end{equation}
arising from the linearizations of $\mathcal E_\omega$ and $\mathcal E_\varphi^{(5)}$ respectively, while for $G_3=\gamma_3\Phi X_\Phi$,
\begin{equation}
\begin{split}
&-4q^5(\lambda q-M)A'' - 4q^4(3\lambda q-2M)A' \\
&\quad - 4q^4(\lambda q-M)W'' - 4\lambda q^4 W' \\
&\quad + 8\lambda^2(\lambda q-M) = 0,
\end{split}
\label{eq:app_Ew3}
\end{equation}
\begin{equation}
\begin{split}
&8q^6(\lambda q-M)A'' + 8q^5(2\lambda q-M)A' \\
&\quad + 4q^5(\lambda q-M)W'' + 4\lambda q^5 W' \\
&\quad + 8\lambda\Big(20M^2 - 31M\lambda q + 10\lambda^2 q^2\Big) = 0.
\end{split}
\label{eq:app_Ep3}
\end{equation}

Equations~\eqref{eq:app_Ew5}--\eqref{eq:app_Ep3} are two coupled second-order ODEs in two unknowns, $A(q)$ and $W(q)$, in each sector.

The key simplification is that, with the normalization above, a suitable linear combination of $E_\varphi^{(1)}$ and $E_\omega^{(1)}$ (where $E_\varphi^{(1)}$ and $E_\omega^{(1)}$ denote, respectively, the left-hand sides of Eqs.~\eqref{eq:app_Ep5}/\eqref{eq:app_Ep3} and~\eqref{eq:app_Ew5}/\eqref{eq:app_Ew3}) eliminates \emph{both} $W''$ and $W'$ simultaneously. In the $G_5$ sector the $W$-terms of Eqs.~\eqref{eq:app_Ew5} and~\eqref{eq:app_Ep5} are equal and opposite, so the required combination is $E_\varphi^{(1)}+E_\omega^{(1)}$; in the $G_3$ sector the $W$-terms of Eq.~\eqref{eq:app_Ep3} carry one extra power of $q$ relative to those of Eq.~\eqref{eq:app_Ew3}, so the required combination is instead $E_\varphi^{(1)}+q\,E_\omega^{(1)}$. Dividing the resulting single equation by the common overall factor of $4q^5$ ($G_5$ sector) or $4q^6$ ($G_3$ sector) collapses the system to the single master equation for $A(q)$ quoted in the main text,
\begin{equation}
(\lambda q-M)A_i''(q)+\lambda A_i'(q)=\mathcal{S}_i(q),
\label{eq:master}
\end{equation}
with sources
\begin{equation}
\begin{aligned}
\mathcal{S}_5(q) &= \frac{2}{\Lambda^2 q^5} \bigl(8\Lambda^2M\lambda^2q^2 - 2\Lambda^2\lambda^3q^3 - 2M + \lambda q\bigr), \\
\mathcal{S}_3(q) &= -\frac{2\lambda(10M-11\lambda q)(2M-\lambda q)}{q^6}\,.
\end{aligned}
\label{eq:S_exact}
\end{equation}
This decoupling is the 2-D analogue of the familiar trick in perturbed BH spacetimes whereby a suitable combination of the Hamiltonian and momentum constraints yields a single master equation for the physical (gauge-invariant) perturbation.

\begin{table}[t]
\centering
\renewcommand{\arraystretch}{1.5}
\begin{tabular}{lcc}
\toprule
Sector & $A(q\to\infty)$ & $A(q\to M/\lambda)$ \\
\midrule
$G_3=\gamma_3\Phi X$ & $-\dfrac{22\lambda^2}{9q^3}$ & $-\dfrac{2\lambda^5}{3M^3}\ln\epsilon$ \\[8pt]
$G_5=\gamma_5\Phi$ & $-\dfrac{4\lambda^2}{q}$ & $\Big(\dfrac{\lambda^3}{3\Lambda^2M^3}-\dfrac{4\lambda^3}{M}\Big)\ln\epsilon$ \\
\bottomrule
\end{tabular}
\caption{Leading far-field and near-horizon behavior of the master-equation solution $A(q)$ in each Horndeski sector, with $\epsilon\equiv q-M/\lambda\to0^+$. In both sectors $A\to0$ at infinity, so $\alpha\to1$, while the horizon limit is logarithmically divergent, foreshadowing the coordinate resolution of Appendix~\ref{app:shift}.}
\label{tab:app_asymptotics}
\end{table}

\subsection{Boundary conditions and closed-form profiles}
\label{app:closed_form}

The homogeneous version of the master equation~\eqref{eq:master}, $(\lambda q-M)A''+\lambda A'=0$, is first order in $A'$ and integrates immediately: writing $B\equiv A'$, $(\lambda q-M)B'=-\lambda B$ gives $B\propto(\lambda q-M)^{-1}$, so that
\begin{equation}
A_{\rm hom}(q) = C_1 + C_2\,\ln(\lambda q-M),
\label{eq:app_Ahom}
\end{equation}
for constants $C_1,C_2$. Since $\ln(\lambda q-M)\to+\infty$ as $q\to\infty$, demanding asymptotic flatness ($A\to0$ at spatial infinity) forces $C_2=0$; the remaining constant $C_1$ merely rescales $\alpha$ by a constant amount, and requiring $\alpha\to1$ at infinity (i.e.\ no residual finite renormalization of the dilaton at spatial infinity) fixes $C_1=0$ as well. Both integration constants of the homogeneous problem are therefore fixed uniquely by the physical boundary conditions at $q\to\infty$. With no freedom left over, the inhomogeneous solution of Eq.~\eqref{eq:master} is unique.

The inhomogeneous solution is obtained by variation of parameters. With integrating factor $\mu(q)=\lambda q-M$, Eq.~\eqref{eq:master} may be written as the exact derivative
\begin{equation}
\frac{d}{dq}\Big[(\lambda q-M)\,A_i'(q)\Big] = \mathcal S_i(q),
\label{eq:app_exact_form}
\end{equation}
which integrates once to give $(\lambda q-M)A'(q)$ as an indefinite integral of $\mathcal S_i(q)$, and a second time (dividing by $\lambda q - M$ and integrating again) to give $A(q)$ itself, with both integration constants fixed to zero by the argument above. The profile $W(q)$ is then recovered algebraically: substituting the now-known $A(q)$ back into the linear equation~\eqref{eq:app_Ew5} or~\eqref{eq:app_Ew3} for $W''$ and $W'$ and integrating once more (again fixing the homogeneous constant by $W\to0$ at infinity) yields $W(q)$. Carrying out this double quadrature explicitly for the two sectors, and verifying \emph{a posteriori} by direct substitution back into both members of Eqs.~\eqref{eq:app_Ew5}--\eqref{eq:app_Ep3} (respectively Eqs.~\eqref{eq:app_Ew3}--\eqref{eq:app_Ep3}), gives the closed-form solutions. Defining the bounded combination $\mathcal L(q)\equiv\ln\!\big(\lambda q/(\lambda q-M)\big)=-\ln(1-M/\lambda q)\geq0$, which vanishes at spatial infinity and diverges logarithmically at the horizon, the $G_3=\gamma_3\Phi X$ sector profiles are
\begin{equation}
\begin{aligned}
A_3(q) &= \frac{\lambda}{3M^3q^4}\Bigl[ M\bigl(6M^3-8M^2\lambda q - M\lambda^2q^2-2\lambda^3q^3\bigr) + \\ &2\lambda^4q^4\,\mathcal{L}(q)\Bigr], \\
W_3(q) &= -\frac{2M\lambda}{q^3} + \frac{8\lambda^2}{3q^2} + \frac{2\lambda^3}{3Mq} - \frac{2\lambda^4}{3M^2}\,\mathcal{L}(q),
\end{aligned}
\label{eq:AW3}
\end{equation}
and the $G_5=\gamma_5\Phi$ sector profiles are
\begin{equation}
\begin{aligned}
A_5(q) &= \frac{M\bigl[2M^2+M\lambda q - 2\lambda^2q^2(24\Lambda^2M^2-1)\bigr]}{6\Lambda^2M^3q^3} + \\
&\frac{(12\Lambda^2M^2-1)\lambda^3}{3\Lambda^2M^3}\,\mathcal{L}(q), \\
W_5(q) &= \frac{4M\lambda}{q} - \frac{1}{3\Lambda^2q^2} - \frac{\lambda}{3\Lambda^2Mq} + \frac{\lambda^2}{3\Lambda^2M^2}\,\mathcal{L}(q).
\end{aligned}
\label{eq:AW5}
\end{equation}
Both sectors exhibit logarithmic divergences as $q\to M/\lambda$, confirming the strong-coupling breakdown of the semiclassical perturbation at the classical horizon.

Expanding $\mathcal L(q)=M/\lambda q+M^2/2\lambda^2q^2+\cdots$ at large $q$ and $\mathcal L(q)\sim-\ln\epsilon+\text{const}$ near $\epsilon\equiv q-M/\lambda\to0^+$, Eqs.~\eqref{eq:AW3}--\eqref{eq:AW5} give the leading behaviors summarized in Table~\ref{tab:app_asymptotics}. Every profile vanishes at spatial infinity, confirming that the Horndeski corrections do not disturb the asymptotically flat, linear-dilaton region, while every profile diverges only logarithmically at the horizon, which is precisely the mild, resummable divergence resolved in Appendix~\ref{app:shift}.

\section{Exact 2D Ricci Scalar and Horizon Regularity}
\label{app:curvature}

This appendix gives the derivation behind the curvature results quoted in Sec.~IV. It shows, order by order in $\gamma_i$, that the logarithmic divergences of $A(q)$ and $W(q)$ found in Appendix~\ref{app:closed_form} never appear in a genuine curvature invariant.
To definitively prove that the logarithmic divergences in the fields $A(q)$ and $W(q)$ are coordinate artifacts, we evaluate the exact 2-D Ricci scalar. In the conformal gauge $ds^2=-e^{2\omega}dx^+dx^-$, the curvature is $R=8v\,\omega_{+-}$ and we perform a perturbative expansion $R=R^{(0)}+\gamma_i R^{(1)}+\mathcal{O}(\gamma_i^2)$. On the static slice of Appendix~\ref{app:static_q}, $\omega_{+-}=\omega'+\chi\omega''$ with $\omega=-\tfrac12\ln v$; evaluating this on the unperturbed background $v_0=q$ using $d/d\chi=-\lambda^2\,d/dq$ from Eq.~\eqref{eq:app_qdef} gives $R^{(0)}(q)=8q\,[\omega_0'+\chi\omega_0'']=4M\lambda/q$, reproducing the standard CGHS result.
The background CGHS curvature $R^{(0)}(q)=4M\lambda/q$ yields the standard finite horizon value $R^{(0)}\big|_{\rm hor}=4\lambda^2$. Substituting the closed-form solutions $W_i(q)$ into the curvature definition, the near-horizon expansion ($\epsilon\equiv q-M/\lambda\to0^+$) gives
\begin{equation}
\begin{aligned}
R^{(1)}\big|_{G_3} &= -\frac{8\lambda^7}{M^3} + \epsilon\left[\frac{8\lambda^8}{3M^4}\ln\epsilon + \cdots\right] + \mathcal{O}(\epsilon^2\ln\epsilon), \\
R^{(1)}\big|_{G_5} &= \frac{8\lambda^5}{\Lambda^2M^3} - \frac{16\lambda^5}{M} - \epsilon\left[\frac{4\lambda^6}{3\Lambda^2M^4}\ln\epsilon + \cdots\right] + \\ 
&\mathcal{O}(\epsilon^2\ln\epsilon).
\end{aligned}
\label{eq:R1_expansions}
\end{equation}

Crucially, the coefficients of $\ln(\lambda q-M)$ vanish identically \emph{at} the horizon $\epsilon=0$; the logarithmic terms enter only at $\mathcal{O}(\epsilon)$. This vanishing is a direct consequence of the specific algebraic structure of Eqs.~\eqref{eq:AW3}--\eqref{eq:AW5}: the coefficient of $\mathcal L(q)$ in $W_i(q)$ is a constant, so $R^{(1)}\propto W_i''(q)+\cdots$ picks up a term $\propto \mathcal L''(q)\propto1/(\lambda q-M)^2$ from twice differentiating $\mathcal L(q)=-\ln(1-M/\lambda q)$, which is exactly cancelled at $\mathcal O(\epsilon^0)$ by compensating terms from the rational prefactors of $W_i(q)$; only the subleading $\epsilon\ln\epsilon$ remainder survives, as displayed in Eq.~\eqref{eq:R1_expansions}. This cancellation is the curvature-space manifestation of the coordinate nature of the divergence identified directly in Appendix~\ref{app:shift}. Hence the horizon values are \emph{finite}:
\begin{equation}
\begin{aligned}
R^{(1)}\big|_{\rm hor}^{G_3} &= -\frac{8\lambda^7}{M^3}\,, \quad R^{(1)}\big|_{\rm hor}^{G_5} &= \frac{8\lambda^5}{\Lambda^2M^3}-\frac{16\lambda^5}{M}\,.
\end{aligned}
\label{eq:R1hor}
\end{equation}

To further confirm horizon regularity, we evaluate the covariant gradient invariant $(\nabla R)^2=g^{ab}\partial_a R\partial_b R$. In the static Kruskal conformal frame this evaluates to
\begin{equation}
\begin{aligned}
(\nabla R)^2 &= 4\lambda(\lambda q-M) \Bigl[ q\,\bigl(R^{(0)\prime}\bigr)^2 \\
&\quad + \gamma_i\bigl(W\bigl(R^{(0)\prime}\bigr)^2 + 2q\,R^{(0)\prime}R^{(1)\prime}\bigr) \Bigr] + \mathcal{O}(\gamma_i^2).
\end{aligned}
\label{eq:gradR2}
\end{equation}

The kinematic prefactor $(\lambda q-M)\to0$ at the horizon suppresses the mild $\ln\epsilon$ divergence of $R^{(1)\prime}$, so the invariant vanishes exactly:
\begin{equation}
(\nabla R)^2\big|_{\rm hor}=0 \qquad \text{at } \mathcal{O}(\gamma^0) \text{ and } \mathcal{O}(\gamma_i).
\label{eq:gradR2hor}
\end{equation}
The higher-derivative couplings therefore shift the horizon curvature by a finite $\mathcal{O}(\gamma_i)$ amount without generating any curvature singularity, leaving the event horizon physically smooth. 

Since $R^{(0)}|_{\rm hor}=4\lambda^2$ is finite and nonzero while $R^{(1)}|_{\rm hor}$ is finite by Eq.~\eqref{eq:R1hor}, the ratio
\begin{equation}
\frac{\gamma_i R^{(1)}}{R^{(0)}}\bigg|_{\rm hor} = O(\gamma_i) \ll 1
\label{eq:app_ratio}
\end{equation}
remains parametrically small throughout the near-horizon region for sufficiently small $\gamma_i$. The linear expansion in $\gamma_i$ is controlled exactly where the auxiliary Kruskal fields $A,W$ individually diverge, because the divergence never propagates into the physical invariant $R$.

As an independent verification of Eq.~\eqref{eq:R1hor}, the near-horizon region can also be reached via a tortoise-gauge Frobenius (power-series) expansion of the field equations directly in the variable $t=e^{2\lambda x}$, rather than through the closed-form solution of Appendix~\ref{app:closed_form}. In that expansion the perturbed dilaton and conformal factor are written as power series $\psi_i=\sum_n a_nt^n$, $\chi_i=\sum_n c_nt^n$ around the horizon $t\to0$, with the leading coefficients $a_0$ (an undetermined dilaton zero mode) and $a_1$ (a matching parameter) left free by the indicial structure of the recursion, and all higher $a_n,c_n$ fixed algebraically in terms of $a_1$, $M$, and $\lambda$. Reconstructing $R^{(1)}|_{\rm hor}$ from this series gives an expression of the form $R^{(1)}|_{\rm hor}\propto M c_1+\lambda c_0$; direct substitution of the recursion coefficients shows that the dependence on the free matching parameter $a_1$ cancels \emph{identically} between $c_0$ and $c_1$, leaving a unique, $a_1$-independent horizon curvature that agrees with Eq.~\eqref{eq:R1hor}. This cross-check, performed in a completely different coordinate patch and expansion scheme from the static-Kruskal construction of Appendix~\ref{app:static_q}, confirms that Eq.~\eqref{eq:R1hor} is a genuine, gauge-independent statement about the physical horizon curvature and not an artifact of the particular closed-form solution used to derive it.

\section{Horizon Regularization and the Interior Singularity}
\label{app:shift}

Appendix~\ref{app:curvature} showed that the scalar curvature $R$ is finite at the horizon, despite the logarithmic divergences in the auxiliary Kruskal-frame functions $A(q)$ and $W(q)$. Section~\ref{sec:remnant_main} demonstrated that this coordinate artifact is resolved via the shifted Kruskal frame $X^\pm = (Y^\pm)^{1+s_i}$, which resums the perturbed lapse function. This appendix provides the technical evaluation of the invariant limits that define the shifted surface gravity and analyzes the unresolved interior singularity.

\subsection{The logarithm divergence of the dilaton}

Since $u=e^{-2\varphi}$ is a genuine spacetime scalar as it is related to the dilaton and is not an auxiliary coordinate-dependent combination, like for instance, $A(q)$ and $W(q)$ individually. Rewriting $u=q+\gamma_i u_1(q)$ with $u_1\equiv Aq+W$, substituting the exact closed-form solutions~\eqref{eq:AW3}--\eqref{eq:AW5} and expanding near $\epsilon\equiv q-M/\lambda\to0^+$ gives:
\begin{align}
u_1\big|_{G_3}\xrightarrow{\ \rm hor\ }-\frac{\lambda^4}{3M^2}\quad(\text{finite}), \\
\qquad
u_1\big|_{G_5}\xrightarrow{\ \rm hor\ }-4\lambda^2\ln\epsilon\quad(\text{divergent}).
\label{eq:u1hor}
\end{align}
For the $G_3$ sector, the individual logarithms in {$q \, A_3(q)$} and $W_3(q)$ cancel exactly in the invariant combination $u_1=Aq+W$. Thus, the divergence in the $G_3$-sector Kruskal profiles was entirely an artifact of splitting the scalar $u$ into the two separately log-divergent pieces $\alpha$ and $v$, and never afflicted the physical dilaton. 

In contrast for the $G_5$ sector, a genuine logarithm survives in the scalar $u$ itself. Since Appendix~\ref{app:curvature} already established that the curvature invariant $R$ is finite at the horizon in \emph{both} sectors, this surviving $G_5$ logarithm cannot signal a curvature singularity, instead it is a defect of the coordinate chart itself, i.e.\ the original Kruskal chart $(X^+,X^-)$ degenerates at $q=M/\lambda$ even though the underlying geometry does not.

\subsection{The shifted Kruskal frame}
\label{app:shift_construction}

Outside the horizon, $X^+>0$, $X^-<0$, so $\chi=X^+X^-<0$ and $\epsilon=-\lambda^2\chi>0$, the metric is $ds^2=-\Omega\,dX^+dX^-$ with conformal factor $\Omega=e^{2\omega}=1/v$. Near the horizon, $v=q+\gamma_iW=M/\lambda+\epsilon+\gamma_i(w_c\ln\epsilon+w_0)+\cdots$, where $w_c$ is the coefficient of $\ln\epsilon$ in $W_i(q)$ (read off from Eq.~\eqref{eq:AW3} or~\eqref{eq:AW5}). Hence,
\begin{equation}
\Omega = \frac1v = \frac{\lambda}{M} - \gamma_i\,\frac{\lambda^2}{M^2}\,w_c\,\ln\epsilon + (\text{regular}) + \cdots.
\label{eq:app_Omega_log}
\end{equation}
At $\gamma_i=0$ the background factor $\Omega_0\to\lambda/M$ is manifestly finite. The unperturbed Kruskal chart is regular at the horizon, and the single concerning piece is the $\gamma_i w_c\ln\epsilon$ term. Using $\epsilon=-\lambda^2\chi=-\lambda^2X^+X^-$,
\begin{equation}
\ln\epsilon = \ln X^+ + \ln(-X^-) + \text{const},
\end{equation}
so the divergence splits \emph{symmetrically} between the two null directions, which is precisely the structure needed for it to be absorbed by a factorized reparametrization $X^+=X^+(Y^+)$, $X^-=X^-(Y^-)$ that treats each null coordinate independently.

Define the shifted Kruskal coordinates $(Y^+,Y^-)$ by the power-law map
\begin{equation}
X^+ = (Y^+)^{1+s_i}, \quad -X^- = (-Y^-)^{1+s_i}, \quad s_i \equiv \gamma_i\,\frac{\lambda}{M}\,w_c.
\label{eq:app_Ytransf}
\end{equation}
This reduces to the identity at $\gamma_i=0$ and differs from it only at $O(\gamma_i)$, so it is a small, near-horizon coordinate shift rather than a global redefinition. Under Eq.~\eqref{eq:app_Ytransf} the Kruskal product transforms as $\epsilon=\lambda^{-2s_i}\epsilon_Y^{1+s_i}$ with $\epsilon_Y\equiv-\lambda^2Y^+Y^-$, and the Jacobian of the transformation contributes
\begin{equation}
\frac{\partial X^+}{\partial Y^+}\frac{\partial X^-}{\partial Y^-}
= (1+s_i)^2\Big(\frac{\epsilon_Y}{\lambda^2}\Big)^{s_i}
= 1 + s_i\ln\epsilon_Y + \text{const} + O(\gamma_i^2).
\end{equation}
{Multiplying Eq.~\eqref{eq:app_Omega_log} by this Jacobian, we evaluate the conformal factor in the new chart. Note that because $s_i \propto \gamma_i$, the exact logarithmic mapping $\ln\epsilon = (1+s_i)\ln\epsilon_Y + \text{const}$ reduces to $\ln\epsilon = \ln\epsilon_Y + \text{const} + O(\gamma_i^2)$ when inserted into the $O(\gamma_i)$ perturbed term. Dropping all second-order corrections, the shifted conformal factor becomes:}
\begin{align}
\Omega_Y &= \Omega,\frac{\partial X^+}{\partial Y^+}\frac{\partial X^-}{\partial Y^-} \\
&= \frac{\lambda}{M} + \Big[\frac{\lambda}{M}s_i - \gamma_i\frac{\lambda^2}{M^2}w_c\Big]\ln\epsilon_Y + O(\gamma_i^2). \nonumber 
\end{align}
The coefficient of $\ln\epsilon_Y$ vanishes \emph{exactly} for the choice $s_i=\gamma_i(\lambda/M)w_c$ already made in Eq.~\eqref{eq:app_Ytransf}: the shifted conformal factor $\Omega_Y$ is completely regular at the new horizon $Y^+Y^-=0$. The $G_5$-sector logarithm was therefore a coordinate artifact exactly as anticipated from the finiteness of $R$.

Near the horizon, the lapse behaves as
\begin{align}
f &= \frac{\epsilon}{v} = \frac{\lambda}{M}\,\epsilon\Big[1-\gamma_i\frac{\lambda}{M}w_c\ln\epsilon\Big] = \frac{\lambda}{M}\,\epsilon^{1+\delta_i}+O(\gamma_i^2), \nonumber \\
\delta_i &\equiv -\gamma_i\frac{\lambda}{M}w_c = -s_i.
\label{eq:app_lapse_resum}
\end{align}
A simple zero $f\propto\epsilon$ has been deformed into the modified power law $f\propto\epsilon^{1+\delta_i}$. Since the Kruskal coordinates satisfy $X^\pm\sim\pm e^{\pm\kappa r_*}$ in terms of the tortoise coordinate $r_*=\int d\epsilon/f$, a shift $\kappa\to\kappa(1+\delta_i)$ of the near-horizon exponent is the power-law reparametrization of Eq.~\eqref{eq:app_Ytransf}. In other words, Eq.~\eqref{eq:app_Ytransf} is nothing but the passage from the \emph{background}-surface-gravity Kruskal frame to the \emph{corrected}-surface-gravity one, it relocates the horizon to its true position, and $\delta_i$ is the fractional shift of the surface gravity computed independently in the next subsection. The linear $\ln\epsilon$ divergence found throughout Appendices~\ref{app:eom} and~\ref{app:curvature} is simply the first-order Taylor term of this resummed power $\epsilon^{1+\delta_i}$; working directly in $(Y^+,Y^-)$ removes the apparent breakdown at any finite order.

\subsection{Hawking temperature of the perturbed black hole}
\label{app:temperature}

The surface gravity is defined invariantly from the norm of the horizon Killing vector $\xi=\lambda(X^+\partial_+-X^-\partial_-)$, whose norm-squared is $|\xi|^2=-f$ with $f=-\lambda^2\chi/v$. For a function of $\chi$ alone, $(\nabla f)^2=-4v\chi(f')^2$ (with $'=d/d\chi$), so
\begin{equation}
\kappa^2 = \lim_{\rm hor}\frac{(\nabla f)^2}{4f}
\Longrightarrow
\kappa = \lim_{\rm hor}\frac{|v\,\partial_\chi f|}{\lambda} = \lambda\Big|1-\lim_{\rm hor}\frac{\chi\,v'}{v}\Big|,
\label{eq:app_kappadef}
\end{equation}
which reproduces $\kappa_0=\lambda$ (i.e.\ $T_H=\lambda/2\pi$) on the unperturbed CGHS background. Writing $v(\chi)=q(\chi)+\gamma_iW(q(\chi))$ with $q=M/\lambda-\lambda^2\chi$ and expanding $\chi v'/v$ to $O(\gamma_i)$ using $\lambda^2\chi=-(\lambda q-M)/\lambda$, the terms proportional to $(\lambda q-M)/\lambda$ and to $(\lambda q-M)W(q)/\lambda q^2$ both vanish at the horizon ($\epsilon\ln\epsilon\to0$ as $\epsilon\to0^+$, so even the divergent piece of $W(q)$ is harmless here), leaving only the term built from $W'(q)$:
\begin{equation}
\lim_{\rm hor}\frac{\chi v'}{v} = \frac{\gamma_i}{M}\lim_{q\to M/\lambda}(\lambda q-M)\,W_i'(q).
\label{eq:app_chivprime_v}
\end{equation}
The two limits appearing here were already isolated in the closed-form solutions of Appendix~\ref{app:closed_form} and evaluate, by direct differentiation of Eqs.~\eqref{eq:AW3}--\eqref{eq:AW5}, to
\begin{align}
\lim_{q\to M/\lambda}(\lambda q-M)\,W_3'(q)=\frac{2\lambda^5}{3M^2}, \\
\qquad
\lim_{q\to M/\lambda}(\lambda q-M)\,W_5'(q)=-\frac{\lambda^3}{3\Lambda^2M^2}.
\label{eq:app_Wprime_limits}
\end{align}
Substituting Eq.~\eqref{eq:app_Wprime_limits} into Eqs.~\eqref{eq:app_chivprime_v}--\eqref{eq:app_kappadef}, the corrected surface gravities are
\begin{equation}
\kappa_{G_3}=\lambda\Big[1-\frac{2\gamma_3\lambda^5}{3M^3}\Big],
\qquad
\kappa_{G_5}=\lambda\Big[1+\frac{\gamma_5\lambda^3}{3\Lambda^2M^3}\Big],
\label{eq:app_kappa_final}
\end{equation}
so that the fractional Hawking-temperature shifts $T_H=\kappa/2\pi$ are
\begin{equation}
\frac{\delta T_H}{T_H}\bigg|_{G_3}=-\frac{2\gamma_3\lambda^5}{3M^3},
\qquad
\frac{\delta T_H}{T_H}\bigg|_{G_5}=+\frac{\gamma_5\lambda^3}{3\Lambda^2M^3}.
\label{eq:app_dT}
\end{equation}
One checks directly that $\delta T_H/T_H=\delta_i$ of Eq.~\eqref{eq:app_lapse_resum}; the horizon-coordinate shift of Appendix~\ref{app:shift_construction} and the temperature shift computed here are the same effect viewed two ways, with the original logarithm being the linearization of the resummed near-horizon power law. For $\gamma_3>0$ the $G_3$ coupling \emph{lowers} the surface gravity (the BH becomes colder as it shrinks, with the effect $\propto(\lambda/M)^3\lambda^2$ largest for light holes), while for $\gamma_5>0$ the $G_5$ coupling makes the BH \emph{hotter}, controlled by the dimensionless ratio $\lambda^3/\Lambda^2M^3$. Equation~\eqref{eq:app_dT} for the $G_3$ sector is the exact linear-order statement that the main text's closed-form expression $T_H(M)=(\lambda/2\pi)(1-M_{\rm r}^3/M^3)$, with $M_{\rm r}\equiv(2\gamma_3\lambda^5/3)^{1/3}$, resums; as emphasized in the main text, this resummation is a physically motivated extrapolation; it is under full perturbative control for $M\gg M_{\rm r}$ but should not be taken as an exact non-perturbative statement all the way down to $M\to M_{\rm r}$, since the perturbative parameter $\gamma_3\lambda^5/M^3$ itself becomes $O(1)$ there.

\subsection{Nature of the interior singularity}
\label{app:interior}

The classical CGHS BH possesses a curvature singularity in its interior at $u=e^{-2\varphi}\to0$, i.e.\ $q\to0^+$, where $R^{(0)}=4M\lambda/q\to+\infty$. Evaluating the Horndeski curvature corrections in the same limit gives
\begin{equation}
R^{(1)}\big|_{G_3}=\frac{152\,M^2\lambda^2}{q^5}+\cdots,
\qquad
R^{(1)}\big|_{G_5}=\frac{44\,M\lambda}{3\Lambda^2 q^4}+\cdots,
\label{eq:app_R1sing}
\end{equation}
The corrections diverge with a \emph{higher} power of $1/q$ than the background curvature.  The relative ratios are:
\begin{equation}
\!\!\!\!\! \frac{\gamma_3 R^{(1)}}{R^{(0)}}\bigg|_{G_3}=38\,\gamma_3 M\lambda\,q^{-4},~
\frac{\gamma_5 R^{(1)}}{R^{(0)}}\bigg|_{G_5}=\frac{11\,\gamma_5}{3\Lambda^2}\,q^{-3}.
\label{eq:app_ratio_interior}
\end{equation}
Both ratios grow without bound as $q\to0$, demonstrating that the Horndeski corrections fail to resolve the classical interior singularity at the linear level. Setting the ratio in Eq.~\eqref{eq:app_ratio_interior} to unity defines the radius where the perturbative expansion in $\gamma_i$ breaks down:
\begin{equation}
q_*^{G_3}\sim\big(38\,\gamma_3 M\lambda\big)^{1/4},
\qquad
q_*^{G_5}\sim\Big(\frac{11\,\gamma_5}{3\Lambda^2}\Big)^{1/3},
\label{eq:app_qstar}
\end{equation}
matching the estimate $q_*\sim(\gamma_3M\lambda)^{1/4}$ quoted in the main text. Inside $q_*$, the linear description is invalid, necessitating a fully non-perturbative treatment or ultraviolet completion. However, the exterior cold-remnant physics is governed by the near-horizon region $q\to M/\lambda$, far from $q_*$ for small $\gamma_3$, ensuring the validity of the EFT prediction for the remnant state.

\section{Evaporation Law, the Lifetime Integral, and the Cold Remnant}
\label{app:remnant}

This appendix derives the modified evaporation law of Sec.~V, the exact lifetime integral, its closed-form antiderivative, the double-pole argument for an infinite lifetime, and the late-time $1/t$ freeze-out law.

\subsection{Hawking evaporation and the lifetime integral}

A two-dimensional conformal field of central charge $c=1$ radiates with Stefan-Boltzmann luminosity $L=(\pi/12)T_H^2$, so the BH mass obeys $dM/dt=-(\pi/12)T_H^2=-\kappa^2/48\pi$. Specializing to the $G_3$ sector with $\gamma_3>0$, whose temperature $T_H(M)=(\lambda/2\pi)(1-M_{\rm r}^3/M^3)$ was derived in Appendix~\ref{app:shift} and resummed into the closed form quoted in the main text, this becomes
\begin{equation}
\frac{dM}{dt} = -\frac{\lambda^2}{48\pi}\Big(1-\frac{M_{\rm r}^3}{M^3}\Big)^2, \qquad M_{\rm r}\equiv\Big(\frac{2\gamma_3\lambda^5}{3}\Big)^{1/3}.
\label{eq:app_dMdt}
\end{equation}
Separating variables, the time to evaporate from an initial mass $M_0$ down to some $M>M_{\rm r}$ is
\begin{equation}
t(M_0\to M) = \frac{4\pi^2}{k\lambda^2}\int_M^{M_0}\frac{M'^6\,dM'}{(M'^3-M_{\rm r}^3)^2}, \qquad k=\frac{\pi}{12},
\label{eq:app_lifetime_int}
\end{equation}
where the integrand has been simplified using the partial-fraction identity
\begin{equation}
\frac{M^6}{(M^3-M_{\rm r}^3)^2} = 1 + \frac{2M_{\rm r}^3}{M^3-M_{\rm r}^3} + \frac{M_{\rm r}^6}{(M^3-M_{\rm r}^3)^2}.
\label{eq:app_partialfrac}
\end{equation}

\begin{figure}[t]
\centering
\includegraphics[width=\columnwidth]{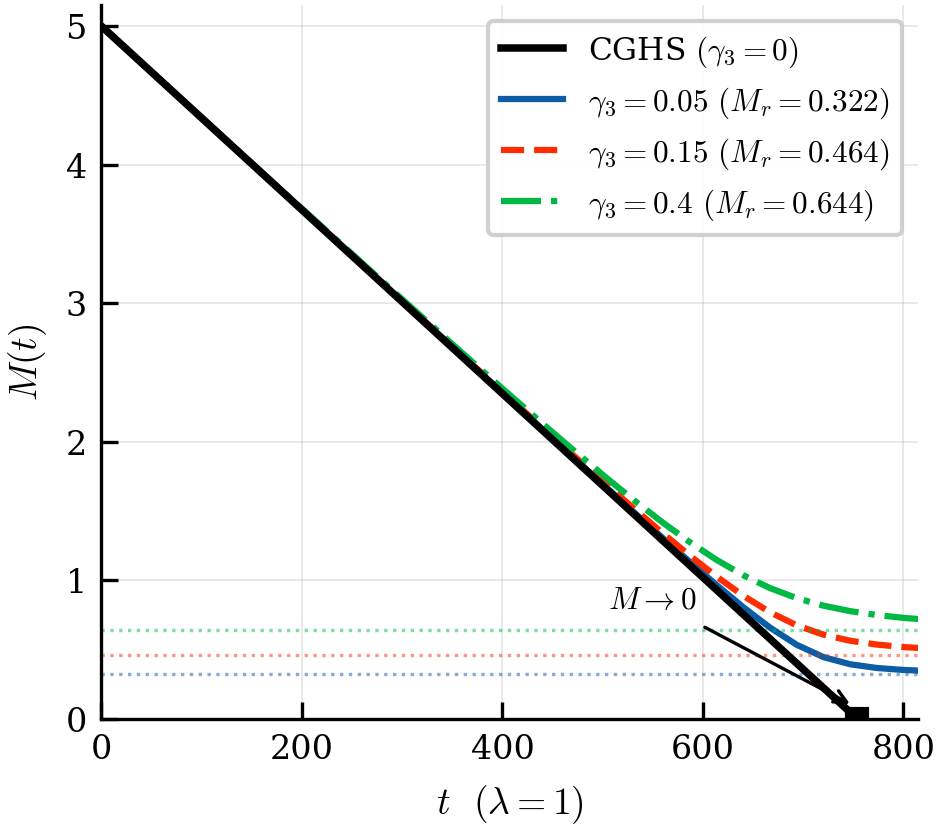}
\caption{Full evaporation history $M(t)$ for the unperturbed CGHS BH ($\gamma_3=0$, solid black), which evaporates linearly to $M=0$ in finite time $t_f\approx754\,\lambda^{-1}$, compared with representative Horndeski-corrected trajectories ($\gamma_3=0.05,0.15,0.4$). The corrected curves decelerate and asymptotically freeze onto their respective cold remnant masses $M_{\rm r}=(2\gamma_3\lambda^5/3)^{1/3}$ (dotted horizontal lines) via the double-pole mechanism of Eq.~\eqref{eq:app_antideriv}, never reaching $M=0$. The late-time power-law tail of Eq.~\eqref{eq:app_freeze} is shown separately in Fig.~\ref{fig:Mvst_latetime} of the main text. All curves use $\lambda=1$, $M_0=5$.}
\label{fig:app_Mvst_evaporation}
\end{figure}

Direct integration of Eq.~\eqref{eq:app_partialfrac} gives the exact closed-form antiderivative
\begin{equation}
\begin{aligned}
\int\frac{M^6\,dM}{(M^3-M_{\rm r}^3)^2}
= {}& M-\frac{M\,M_{\rm r}^3}{3(M^3-M_{\rm r}^3)}
+\frac{4M_{\rm r}}{9}\ln(M-M_{\rm r}) \\
&-\frac{2M_{\rm r}}{9}\ln\!\big(M^2+MM_{\rm r}+M_{\rm r}^2\big) \\
&-\frac{4M_{\rm r}}{3\sqrt3}\arctan\!\frac{2M+M_{\rm r}}{\sqrt3\,M_{\rm r}}.
\end{aligned}
\label{eq:app_antideriv}
\end{equation}
The crucial feature is the double pole at $M=M_{\rm r}$: as $M\to M_{\rm r}^+$, the partial-fraction integrand behaves as $M_{\rm r}^2/[9(M-M_{\rm r})^2]$, whose antiderivative $-M_{\rm r}^2/[9(M-M_{\rm r})]$, visible as the leading singular part of the first term in Eq.~\eqref{eq:app_antideriv} diverges. Hence
\begin{equation}
t(M_0\to M_{\rm r}) = \infty:
\end{equation}
the BH takes infinite coordinate time to reach the remnant mass, never crossing it, exactly as illustrated in Fig.~\ref{fig:app_Mvst_evaporation} below and in Fig.~\ref{fig:Mvst_latetime} of the main text.

\subsection{Late-time power-law freeze-out}

Setting $M=M_{\rm r}+\delta$ with $\delta\to0^+$ and expanding Eq.~\eqref{eq:app_dMdt} to leading order in $\delta$ (using $1-M_{\rm r}^3/M^3\simeq3\delta/M_{\rm r}$) linearizes the evaporation law to the Riccati-type equation
\begin{equation}
\frac{d\delta}{dt} = -\mathcal A\,\delta^2, \qquad \mathcal A = \frac{9c\lambda^2}{(2\pi)^2M_{\rm r}^2} = \frac{3\lambda^2}{16\pi M_{\rm r}^2},
\label{eq:app_riccati}
\end{equation}
whose solution, obtained by separating variables ($d(1/\delta)/dt=\mathcal A$), is the power law
\begin{equation}
M(t)-M_{\rm r} = \delta(t) = \frac{1}{\mathcal A t+\delta_0^{-1}} \xrightarrow{t\to\infty} \frac{16\pi M_{\rm r}^2}{3\lambda^2\,t} \sim \frac1t,
\label{eq:app_freeze}
\end{equation}
which is the late-time law already quoted in the main text and displayed in Fig.~\ref{fig:Mvst_latetime}. Since $T_H\propto\delta\to0$ and $dM/dt\propto\delta^2\to0$ together, both the temperature and the luminosity vanish smoothly in this limit, and the evaporation halts dynamically.

\begin{figure}[htbp]
\centering
\includegraphics[width=\columnwidth]{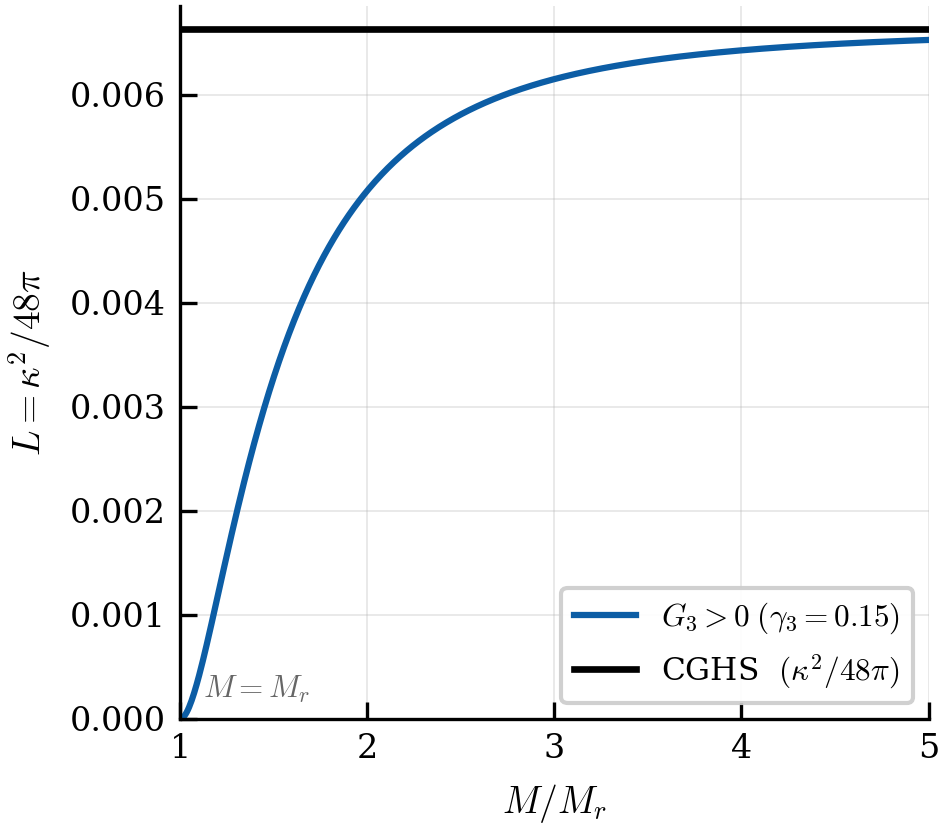}
\caption{Semiclassical luminosity $L=\kappa^2/48\pi$ as a function of the normalized mass $M/M_{\rm r}$ in the $G_3$ sector ($\gamma_3=0.15$). Like the temperature shown in Fig.~\ref{fig:HawTempVsTime} of the main text, $L$ vanishes continuously at $M=M_{\rm r}$ and approaches the constant CGHS value $L^{\rm CGHS}=\lambda^2/48\pi$ (horizontal black line) for $M\gg M_{\rm r}$; since $L\propto\kappa^2\propto\delta^2$ near the remnant while $T_H\propto\kappa\propto\delta$, the luminosity vanishes quadratically faster than the temperature, consistent with the Riccati equation~\eqref{eq:app_riccati} governing $\delta(t)$.}
\label{fig:app_L_vs_MMr}
\end{figure}

The remnant mass scales as $M_{\rm r}\sim\lambda(\gamma_3\lambda^2)^{1/3}$, which is parametrically small but not parametrically separated from the perturbative breakdown scale. The dimensionless expansion parameter controlling Eq.~\eqref{eq:app_dMdt}, namely $\gamma_3\lambda^5/M^3$, is by construction $O(1)$ exactly at $M=M_{\rm r}$. The closed-form evaporation law~\eqref{eq:app_dMdt} and the resulting lifetime formula~\eqref{eq:app_antideriv} should therefore be understood as the physically motivated resummation of the controlled linear-order result of Appendix~\ref{app:shift} rather than as an exact non-perturbative statement. Qualitatively, $T_H(M)$ is monotonically decreasing in $M_{\rm r}^3/M^3$ for $\gamma_3>0$, so the evaporation necessarily slows down and the lifetime integral~\eqref{eq:app_lifetime_int} necessarily develops a pole at some finite mass of order $M_{\rm r}$, even before the precise numerical coefficients of Eq.~\eqref{eq:app_antideriv} are trusted at the percent level.

\section{Conformal Anomaly, Stress Tensor, and the Robinson-Wilczek Construction}
\label{app:anomaly}

This appendix derives the semiclassical stress tensor and gravitational anomaly used in Sec.~VI.

\subsection{Conservation equations in null gauge}

In the conformal null gauge $ds^2=-e^{2\omega}dx^+dx^-$, the nonzero metric components are $g_{+-}=-\tfrac12e^{2\omega}$, $g^{+-}=-2e^{-2\omega}$, $\sqrt{-g}=\tfrac12e^{2\omega}$, and the only nonzero Christoffel symbols are $\Gamma^+_{++}=2\partial_+\omega$, $\Gamma^-_{--}=2\partial_-\omega$, $\Gamma^\mu_{\mu\lambda}=\partial_\lambda\ln\sqrt{-g}=2\partial_\lambda\omega$. Writing out $\nabla_\mu T^\mu{}_\nu=0$ for $\nu=\pm$ and stripping an overall common factor gives the two conservation equations
\begin{align}
\partial_-\langle T_{++}\rangle + \partial_+\langle T_{+-}\rangle - 2(\partial_+\omega)\langle T_{+-}\rangle &= 0, \label{eq:app_consplus}\\
\partial_+\langle T_{--}\rangle + \partial_-\langle T_{+-}\rangle - 2(\partial_-\omega)\langle T_{+-}\rangle &= 0. \label{eq:app_consminus}
\end{align}
The off-diagonal component $\langle T_{+-}\rangle$ couples the two chiralities and, unlike $\langle T_{\pm\pm}\rangle$, is completely fixed algebraically by the trace anomaly rather than by an integration constant. Combining $\langle T^\mu{}_\mu\rangle=2g^{+-}\langle T_{+-}\rangle=-4e^{-2\omega}\langle T_{+-}\rangle$ with the Duff result $\langle T^\mu{}_\mu\rangle=R/24\pi$ (see \citep{Duff:1977ay}) and $R=8e^{-2\omega}\partial_+\partial_-\omega$ gives
\begin{equation}
\langle T_{+-}\rangle = -\frac{1}{12\pi}\partial_+\partial_-\omega = -\frac{1}{96\pi}e^{2\omega}R.
\label{eq:app_Tpm}
\end{equation}
Substituting Eq.~\eqref{eq:app_Tpm} into Eqs.~\eqref{eq:app_consplus}--\eqref{eq:app_consminus} turns each right-hand side into an exact derivative; integrating once gives the full stress tensor
\begin{equation}
\langle T_{\pm\pm}\rangle = -\frac{1}{12\pi}\Big[(\partial_\pm\omega)^2-\partial_\pm^2\omega\Big] + t_\pm(x^\pm),
\label{eq:app_Tpp_solved}
\end{equation}
with $t_\pm(x^\pm)$ state-dependent integration functions fixed by the choice of vacuum (the Unruh state, appropriate for an evaporating BH, sets these to reproduce the standard thermal flux at infinity). By construction, Eqs.~\eqref{eq:app_Tpm}--\eqref{eq:app_Tpp_solved} solve the conservation equations identically; the physical content of $\nabla_\mu\langle T^{\mu\nu}\rangle=0$ here is precisely the determination of the tensor up to $t_\pm$.

\subsection{Explicit components on the Horndeski-corrected background}

\begin{figure}[t]
\centering
\includegraphics[width=\columnwidth]{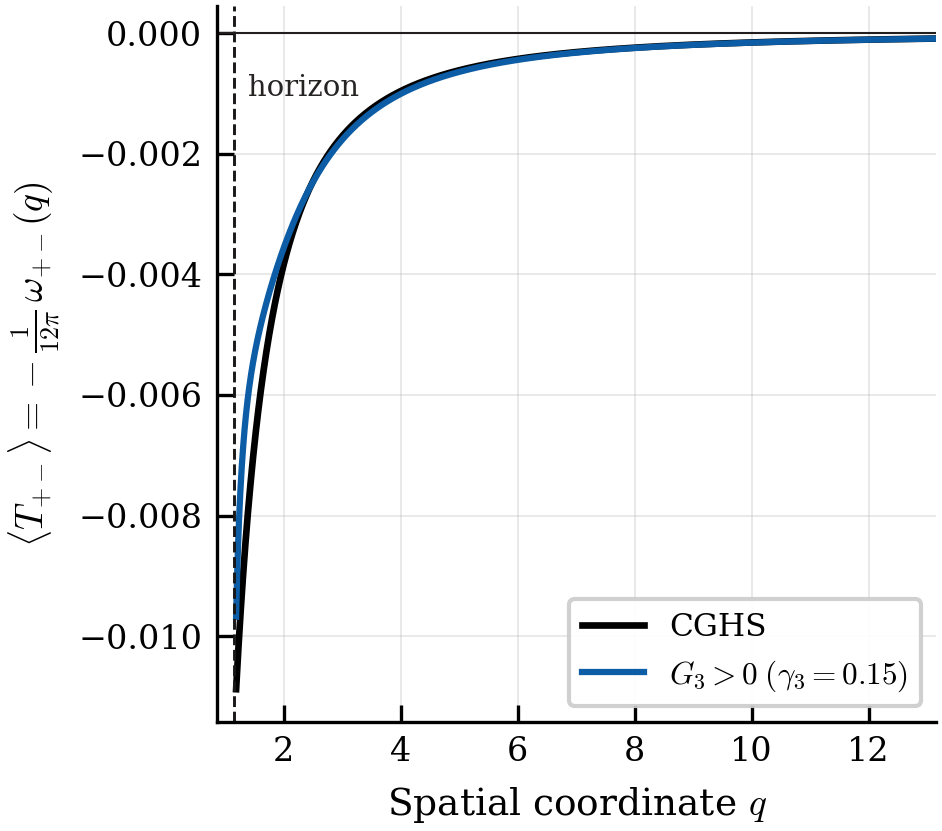}
\caption{Off-diagonal stress-tensor component $\langle T_{+-}\rangle=-\frac{1}{12\pi}\omega_{+-}(q)$ versus the spatial Kruskal coordinate $q$ across the BH exterior ($M=2.5M_{\rm r}$, $\gamma_3=0.15$), fixed entirely by the trace anomaly through Eq.~\eqref{eq:app_Tpm}. As with the geometric component $\langle T_{--}\rangle$ shown in Fig.~\ref{fig:StressTensor} of the main text, the profile vanishes at spatial infinity by Eq.~\eqref{eq:app_Xi_omega_background}; the CGHS background curve remains finite at the horizon, while the $O(\gamma_3)$ correction diverges there as $\epsilon^{-1}$, purely because of the coordinate degeneration discussed above. It never affects a physical invariant.}
\label{fig:app_Tpm_vs_q}
\end{figure}

On the static slice, the geometric (state-independent) content of Eq.~\eqref{eq:app_Tpp_solved} and the trace piece~\eqref{eq:app_Tpm} both reduce to functions of $q$ alone,
\begin{align}
\Xi(q) \equiv (\partial_\chi\omega)^2-\partial_\chi^2\omega = \lambda^4\big[(\partial_q\omega)^2-\partial_q^2\omega\big], \\
\qquad
\omega_{+-}(q) = \lambda^2\Big[-\partial_q\omega+\big(\tfrac{M}{\lambda}-q\big)\partial_q^2\omega\Big],
\label{eq:app_Xi_omega_def}
\end{align}
with $\langle T_{\pm\pm}\rangle=-\tfrac{1}{12\pi}(X^\mp)^2\Xi(q)+t_\pm$ and $\langle T_{+-}\rangle=-\tfrac1{12\pi}\omega_{+-}(q)$, evaluated on $\omega(q)=-\tfrac12\ln v(q)$ with $v=q+\gamma_iW_i(q)$ from Appendix~\ref{app:closed_form}. Direct evaluation on the unperturbed background gives the exact, closed forms
\begin{align}
\Xi_0(q) = -\frac{\lambda^4}{4q^2}, \quad
\omega_{+-}^{(0)}(q) = \frac{M\lambda}{2q^2}, \\
\langle T_{+-}\rangle^{(0)} = -\frac{M\lambda}{24\pi q^2}, \quad
\langle T^\mu{}_\mu\rangle^{(0)} = \frac{M\lambda}{6\pi q} = \frac{R^{(0)}}{24\pi},
\label{eq:app_Xi_omega_background}
\end{align}
consistent with the trace-anomaly relation~\eqref{eq:app_Tpm} applied to the background curvature $R^{(0)}=4M\lambda/q$ of Appendix~\ref{app:curvature}. The $O(\gamma_i)$ corrections $\Xi_1(q)$ and $\omega_{+-}^{(1)}(q)$ follow from Eq.~\eqref{eq:app_Xi_omega_def} by exactly the same substitution used for $R^{(1)}$ in Appendix~\ref{app:curvature}, but unlike the curvature invariant $R$, $\Xi(q)$ and $\omega_{+-}(q)$ are individual tensor components in the Kruskal chart, not diffeomorphism invariants. We have verified that both vanish as $q\to\infty$, so that asymptotic flatness is preserved order by order. At the horizon, by contrast, the background pieces~\eqref{eq:app_Xi_omega_background} are manifestly \emph{finite}; it is the $O(\gamma_i)$ corrections alone that grow without bound as $q\to M/\lambda$, and they do so as inverse powers rather than logarithmically. Because the coefficient of $\mathcal L(q)$ in $W_i(q)$ is a constant, $\Xi_1(q)$ inherits a term $\propto\mathcal L''(q)\propto\epsilon^{-2}$ and $\omega_{+-}^{(1)}(q)$ a term $\propto\epsilon^{-1}$; consequently the $\gamma_i$ expansion of these individual components, unlike that of the invariants of Appendix~\ref{app:curvature}, ceases to be controlled within a narrow strip of the horizon. This is not evidence of any physical singularity, it is precisely the same coordinate degeneration of the baseline Kruskal chart identified and resolved in Appendix~\ref{app:shift}, now appearing in a non-invariant tensor component rather than in the scalar curvature. The explicit profiles of $\langle T_{+-}\rangle$ and the geometric part of $\langle T_{--}\rangle$ across the full exterior region are shown in Fig.~\ref{fig:app_Tpm_vs_q} below and in Fig.~\ref{fig:StressTensor} of the main text.

\subsection{The Robinson-Wilczek construction}

The Robinson-Wilczek derivation~\cite{Robinson:2005pd} obtains the same asymptotic flux from an entirely independent argument, that is, the \emph{gravitational}, rather than conformal, anomaly of the effective chiral theory obtained by integrating out the ingoing modes in a thin shell at the horizon. Writing the geometry in Schwarzschild-type gauge $ds^2=-f(r)\,dt^2+dr^2/f(r)$, the resulting outgoing sector acquires a purely timelike anomalous divergence $\nabla_\mu T^{\chi\mu}{}_\nu=A_\nu=\partial_\mu N^\mu{}_\nu/\sqrt{-g}$, with
\begin{equation}
N^t{}_t=N^r{}_r=0, \quad N^r{}_t = \frac{1}{192\pi}\big(f'^2+f''f\big), \quad {}'\equiv\frac{d}{dr}.
\label{eq:app_Nrt}
\end{equation}
Restoring general covariance requires a compensating flux equal to the value of this anomaly precisely at the horizon, where $f(r_H)=0$ removes the $f''f$ term: $\Phi=N^r{}_t|_{r_H}=f'(r_H)^2/192\pi=\kappa^2/48\pi$ with $\kappa=\tfrac12f'(r_H)$. To evaluate this on the Horndeski-corrected geometry we must construct the tortoise coordinate $r$ explicitly. The exact 2-D identity $R=-f''(r)$, together with the invariant lapse $f(q)=(\lambda q-M)/(\lambda v)$ (normalized to $f\to1$ at infinity) and the curvature $R(q)$ of Appendix~\ref{app:curvature}, fixes the Jacobian $J(q)\equiv dr/dq$ through the first-order linear ODE
\begin{equation}
f_{qq}\,J - f_q\,J_q = -R\,J^3.
\label{eq:app_JODE}
\end{equation}
On the unperturbed background, $J_0=1/(2\lambda q)$, reproducing the familiar relation $q\propto e^{2\lambda r}$; one verifies directly (as we do below for the background piece) that this gives $f_0'(r)=2M/q$ and, via $R^{(0)}=-f_0''(r)$, exactly reproduces $R^{(0)}=4M\lambda/q$. Writing $J=J_0+\gamma_iJ_1$, removing the homogeneous solution $J_1\propto q$ (a pure rescaling of the affine parameter $r$, with no physical content), and evaluating $f'(r)=f_q/J$ at the horizon reproduces exactly the invariant surface gravity already derived independently in Appendix~\ref{app:shift},
\begin{align}
f'(r_H)=2\kappa, \quad \kappa_{G_3}=\lambda\Big[1-\frac{2\gamma_3\lambda^5}{3M^3}\Big], \nonumber \\ \kappa_{G_5}=\lambda\Big[1+\frac{\gamma_5\lambda^3}{3\Lambda^2M^3}\Big].
\end{align}
Since this $\kappa$ was obtained here from the purely geometric tortoise-coordinate construction of Eq.~\eqref{eq:app_JODE}, completely independently of the Killing-vector definition~\eqref{eq:app_kappadef} used in Appendix~\ref{app:shift}. Using $f''=-R$ and $f(r_H)=0$ with $R$ finite at the horizon (Appendix~\ref{app:curvature}), the flux itself is manifestly finite,
\begin{align}
\Phi=N^r{}_t\big|_{r_H}=\frac{\kappa^2}{48\pi}\ \Longrightarrow\
\Phi\big|_{G_3}=\frac{\lambda^2}{48\pi}-\gamma_3\frac{\lambda^7}{36\pi M^3}+O(\gamma_3^2), \nonumber \\
\qquad
\Phi\big|_{G_5}=\frac{\lambda^2}{48\pi}+\gamma_5\frac{\lambda^5}{72\pi\Lambda^2M^3}+O(\gamma_5^2),
\label{eq:app_Phi}
\end{align}
which we have verified, by direct expansion of $\kappa_{G_3}^2/48\pi$ to $O(\gamma_3)$, matches term by term both the conformal-anomaly flux $\langle T_{--}\rangle_\infty$ of the main text and the Stefan-Boltzmann evaporation rate $-dM/dt$ of Eq.~\eqref{eq:app_dMdt}.

\subsection{The anomalous divergence and the background anomaly}

In the gauge $ds^2=-f\,dt^2+dr^2/f$ (where $\sqrt{-g}=1$), the anomaly of Eq.~\eqref{eq:app_Nrt} has only a timelike component,
\begin{align}
A_t = \partial_rN^r{}_t = \frac{1}{192\pi}\partial_r\big(f'^2+f''f\big) \nonumber \\
= -\frac{1}{192\pi}\big(3f'R+f\,\partial_r \, R\big),
\label{eq:app_At_def}
\end{align}
using $f''=-R$. Evaluating Eq.~\eqref{eq:app_At_def} on the unperturbed background, where $f_0(q)=1-M/\lambda q$ and $dq/dr=2\lambda q$ so that $f_0'(r)=2M/q$ and $dR^{(0)}/dr=-8M\lambda^2/q$, gives the exact closed form
\begin{equation}
A_t^{(0)}(q) = \frac{M\lambda(\lambda q-4M)}{24\pi q^2}, \qquad A_t^{(0)}\big|_{\rm hor} = -\frac{\lambda^3}{8\pi},
\label{eq:app_At0}
\end{equation}
which is nonzero and smoothly interpolates from $-\lambda^3/8\pi$ at the horizon to $0$ at spatial infinity. This background piece requires no perturbative expansion and is exact to all orders in $\gamma_i$. The $O(\gamma_i)$ correction $A_t^{(1)}(q)$ is obtained from the same Eq.~\eqref{eq:app_At_def} using the corrected $f_q=f_q^{(0)}+\gamma_iJ$ dependent pieces and $R^{(1)}(q)$ of Appendix~\ref{app:curvature}. Similar to $\Xi_1(q)$ and $\omega_{+-}^{(1)}(q)$ above,  it vanish at spatial infinity (the chiral truncation only matters near the horizon) and is finite at the horizon, since it sources the manifestly finite flux shift already quoted in Eq.~\eqref{eq:app_Phi}. Since $A_t=\partial_rN^r{}_t$ by definition, the fundamental theorem of calculus gives immediately
\begin{equation}
\int_{r_H}^{\infty}A_t\,dr = N^r{}_t\big|_\infty - N^r{}_t\big|_{r_H} = 0-\Phi = -\frac{\kappa^2}{48\pi},
\label{eq:app_anomaly_integral}
\end{equation}
independent of any details of the profile of $A_t(q)$. The energy-momentum that must be added to restore covariance is exactly the flux $\Phi=\kappa^2/48\pi$ found in Eq.~\eqref{eq:app_Phi}, which is the Robinson-Wilczek realization of the cold-remnant evaporation law quoted in the main text and shown, order by order in $M/M_{\rm r}$, in Fig.~\ref{fig:Anomaly}.

\section{Entanglement Entropy, Production Rate and Late-Time Saturation}
\label{app:entropy}

This appendix derives the entropy formulas used in Sec.~VII, the HLW/FPST renormalized entropy, the production-rate ODE system and its closed-form solution $S(M)$, the CGHS limit, the late-time logarithmic law, and the resolution of the apparent entropy/mass-budget tension noted in the main text.

\subsection{The HLW/FPST renormalized entropy}

For a $(1{+}1)$-dimensional conformal field theory in a collapse geometry, the renormalized entropy of the right-moving (outgoing) radiation collected by a distant observer over a retarded-time interval $[u_1,u_2]$ is fixed entirely by the ray-tracing relation $U(u)$, the ingoing Kruskal coordinate evaluated along the outgoing null ray that reaches future null infinity at retarded time $u\equiv x^-=t-x$ (the standard Eddington retarded time, equal to $t-r_*$ in areal gauge)~\cite{Holzhey:1994we}:
\begin{equation}
S_{\rm ren} = \frac{N}{12}\ln\!\left[\frac{\big(U(u_2)-U(u_1)\big)^2}{U'(u_1)U'(u_2)(u_2-u_1)^2}\right],
\label{eq:app_HLW}
\end{equation}
for $N$ free scalar fields (the single-field HLW normalization uses $N=1$; FPST~\cite{Fiola:1994ir} write the same formula with the explicit $N$-dependence since each field contributes to the entanglement independently). The prefactor $N/12$ arises as $c/12$ with central charge $c=N$: a full spatial interval in a CFT contributes $c/3$ to the entanglement entropy, tracing over only the outgoing chiral sector halves this to $c/6$, and the two-endpoint structure of the interval entropy formula contributes a further factor of $\tfrac12$.

For a BH of surface gravity $\kappa$, the late-time ray-tracing relation is the thermal (Schwarzschild-type) trajectory $U(u)=c_1+c_2e^{-\kappa u}$, $U'(u)\propto e^{-\kappa u}$. Substituting into Eq.~\eqref{eq:app_HLW} and taking the interval long compared to $\kappa^{-1}$ reproduces the familiar linear thermal growth $S_{\rm ren}\simeq\tfrac{\kappa}{12}(u_2-u_1)=\tfrac{\pi}{6}T_H(u_2-u_1)$, the $(1{+}1)$-dimensional blackbody entropy at temperature $T_H=\kappa/2\pi$.

\subsection{Entropy production rate and the closed-form \texorpdfstring{$S(M)$}{S(M)}}

For a slowly evaporating (adiabatic) BH the surface gravity drifts with retarded time through the mass, $\kappa=\kappa(M(u))$, and the ray-tracing relation generalizes to $d(\ln U'(u))/du=-\kappa(M(u))$. With the early endpoint $u_1$ held fixed and $u_2=u$ advancing, the growing contribution to Eq.~\eqref{eq:app_HLW} is the redshift of the late endpoint, $-\tfrac1{12}\ln U'(u)$, giving the universal entropy production rate,
\begin{equation}
\frac{dS}{du} = \frac{N}{12}\kappa\big(M(u)\big) = \frac{\pi}{6}NT_H\big(M(u)\big).
\label{eq:app_rate}
\end{equation}
Combining Eq.~\eqref{eq:app_rate} with the (retarded-time) evaporation law $dM/du=-\kappa(M)^2/48\pi$ and $\kappa(M)=\lambda(1-M_{\rm r}^3/M^3)$ gives the coupled system
\begin{equation}
\frac{dM}{du}=-\frac{\kappa(M)^2}{48\pi}, \qquad \frac{dS}{du}=\frac{N}{12}\kappa(M).
\label{eq:app_coupled}
\end{equation}
Dividing the two equations eliminates $u$ and gives the closed relation
\begin{equation}
\frac{dS}{dM} = -\frac{4\pi N}{\kappa(M)} = -\frac{4\pi N}{\lambda}\frac{M^3}{M^3-M_{\rm r}^3},
\label{eq:app_dSdM}
\end{equation}
which integrates from the current mass $M$ up to the initial mass $M_0$ to
\begin{equation}
\begin{split}
S(M) &= \frac{4\pi N}{\lambda}\bigg[(M_0-M) + \frac{M_{\rm r}}{3}\ln\frac{M_0-M_{\rm r}}{M-M_{\rm r}} \\
&\quad -\frac{M_{\rm r}}{6}\ln\frac{M_0^2+M_0M_{\rm r}+M_{\rm r}^2}{M^2+MM_{\rm r}+M_{\rm r}^2} \\
&\quad -\frac{M_{\rm r}}{\sqrt3}\Big(\arctan\frac{2M_0+M_{\rm r}}{\sqrt3M_{\rm r}} - \arctan\frac{2M+M_{\rm r}}{\sqrt3M_{\rm r}}\Big)\bigg].
\end{split}
\label{eq:app_SofM}
\end{equation}
The logarithm $\ln(M-M_{\rm r})$ is the essential new feature relative to CGHS: as $M\to M_{\rm r}^+$ the entropy diverges, but only logarithmically, and as shown in Appendix~\ref{app:remnant} the remnant is reached only after infinite retarded time, so this divergence is never actually attained.

As $M_{\rm r}\to0$, $\kappa\to\lambda$ and Eq.~\eqref{eq:app_SofM} collapses smoothly to $S=(4\pi N/\lambda)(M_0-M)$, linear in the radiated mass, exactly as required. With the unperturbed constant-rate law $dM/du=-\lambda^2/48\pi$, this integrates to $M(u)=M_0-\lambda^2u/48\pi$ and hence $S(u)=(N\lambda/12)u$: the BH evaporates completely in the finite retarded time $u_f=48\pi M_0/\lambda^2$, terminating at the finite entropy
\begin{equation}
S_{\rm CGHS}^{\rm final} = \frac{4\pi NM_0}{\lambda}.
\label{eq:app_Sfinal}
\end{equation}
In the semiclassical approximation, the fine-grained entropy grows monotonically to the end of evaporation, with no Page turnover, signalling information loss.

\subsection{Late-time logarithmic saturation}
Near the remnant, write $M=M_{\rm r}+\epsilon$; then $\kappa(M)\simeq3\lambda\epsilon/M_{\rm r}$ and the mass equation of Eq.~\eqref{eq:app_coupled} becomes the same Riccati-type equation as Eq.~\eqref{eq:app_riccati} of Appendix~\ref{app:remnant} (now in retarded time $u$ rather than coordinate time $t$), with solution
\begin{equation}
\epsilon(u) = \frac{16\pi M_{\rm r}^2}{3\lambda^2(u+C)} \xrightarrow{u\to\infty} \frac{16\pi M_{\rm r}^2}{3\lambda^2u},
\end{equation}
so that the surface gravity falls as
\begin{equation}
\kappa(u) \simeq \frac{3\lambda}{M_{\rm r}}\epsilon(u) = \frac{16\pi M_{\rm r}}{\lambda u}.
\end{equation}
The entropy production rate of Eq.~\eqref{eq:app_rate} then falls as
\begin{equation}
\frac{dS}{du} = \frac{N}{12}\kappa(u) \simeq \frac{4\pi NM_{\rm r}}{3\lambda u},
\end{equation}
which integrates once more to the explicit late-time entropy quoted in the main text,
\begin{equation}
S(u) \simeq S_{\rm sat} + \frac{4\pi NM_{\rm r}}{3\lambda}\ln u.
\label{eq:app_lnu}
\end{equation}
The entropy transitions from linear growth to this asymptotic logarithmic law rather than saturating at a finite value. Since the production rate vanishes only as $u\to\infty$ and never identically, $S(u)$ has no finite maximum, but it also never returns to a genuine Page-curve-like decrease. The entanglement is permanently trapped in the long-lived cold remnant rather than being returned to the outgoing radiation field.

\subsection{Resolving the entropy/mass-budget tension}

At first sight it seems paradoxical that the Horndeski BH, which radiates \emph{less} total mass than the CGHS BH (it retains the finite remnant mass $M_{\rm r}$ rather than radiating all the way to $M=0$), can nonetheless produce \emph{more} total entanglement entropy over the course of its evolution, as illustrated by the Horndeski curves in Fig.~\ref{fig:entropy}(a) eventually overtaking the frozen CGHS final value. Equation~\eqref{eq:app_dSdM} resolves this directly, the entropy produced \emph{per unit mass radiated} is
\begin{equation}
\frac{dS}{|dM|} = \frac{4\pi N}{\kappa(M)},
\label{eq:app_efficiency}
\end{equation}
which is simply $dS=|dM|/T_H$, the ordinary thermodynamic entropy released by a body radiating at temperature $T_H$. Because $\kappa(M)=\lambda(1-M_{\rm r}^3/M^3)<\lambda$ at every mass $M>M_{\rm r}$, the efficiency~\eqref{eq:app_efficiency} \emph{exceeds} the constant CGHS value $4\pi N/\lambda$ throughout the entire evaporation history, and diverges as $M\to M_{\rm r}^+$. The Horndeski BH is therefore a strictly less efficient radiator of mass but a strictly more efficient producer of entropy per unit mass at every stage; sufficiently close to the remnant this enhancement is unbounded, since a body cooling toward absolute zero necessarily produces divergent entropy per unit energy released ($dS=|dM|/T_H\to\infty$ as $T_H\to0$, even though the total available energy $M_0-M_{\rm r}$ remains finite). This is the same physics, viewed thermodynamically, as the logarithmic divergence of Eq.~\eqref{eq:app_SofM} at $M\to M_{\rm r}^+$, an inefficient but eternal radiator can out-produce an efficient but short-lived one in total entanglement, without ever violating the finite mass budget $M_0-M_{\rm r}$ available to it.

%

\end{document}